\documentclass[twocolumn]{aastex701}

\usepackage{xcolor}
\shorttitle{Evaluating extremely metal-poor candidates from Gaia XP}
\shortauthors{Thai et al.}

\usepackage{booktabs}
\usepackage[bottom]{footmisc}
\usepackage{float}
\graphicspath{{./}}
\newcommand{\teff}{T_\mathrm{eff}}
\newcommand{\logg}{\log(g)}
\newcommand{\vt}{\nu_t}
\newcommand{\feh}{[\mathrm{Fe/H}]}
\newcommand{\mh}{[\mathrm{M/H}]}
\newcommand{\afe}{[\mathrm{\alpha/Fe}]}
\newcommand{\mgfe}{\([\mathrm{Mg/Fe}]\)}
\newcommand{\cafe}{\([\mathrm{Ca/Fe}]\)}
\newcommand{\sife}{\([\mathrm{Si/Fe}]\)}
\newcommand{\tife}{\([\mathrm{Ti/Fe}]\)}
\hypersetup{
	colorlinks=true,
	linkcolor=blue,
	citecolor=blue,
	urlcolor=cyan,
  pdftitle={Evaluating extremely metal-poor candidates from Gaia XP}
}

\begin{document}

\title{Evaluating classifications of extremely metal-poor candidates selected from Gaia XP spectra}

\author[0009-0000-9368-0006]{Riley Thai}
\affiliation{School of Physics \& Astronomy, Monash University, Melbourne, VIC 3800, Australia}
\email[show]{riley.thai@monash.edu}

\author[0000-0003-0174-0564]{Andrew R. Casey}
\affiliation{School of Physics \& Astronomy, Monash University, Melbourne, VIC 3800, Australia}
\affiliation{Center for Computational Astrophysics, Flatiron Institute, 162 5th Avenue, New York, NY 10010, USA}
\email{andrew.casey@monash.edu}

\author[0000-0003-0174-0564]{Alexander P. Ji}
\affiliation{Department of Astronomy \& Astrophysics, University of Chicago, 5640 South Ellis Avenue, Chicago, IL 60637, USA;}
\affiliation{Kavli Institute for Cosmological Physics, University of Chicago, Chicago, IL 60637, USA}
\affiliation{NSF-Simons AI Institute for the Sky (SkAI), 172 E. Chestnut St., Chicago, IL 60611, USA}
\email{alex.ji@uchicago.edu}

\author[0000-0002-0572-8012]{Vedant Chandra}
\affiliation{Center for Astrophysics, Harvard \& Smithsonian, 60 Garden Street, Cambridge, MA 02138, USA}
\email{vedant.chandra@cfa.harvard.edu}

\author[0000-0003-4996-9069]{Hans-Walter Rix}
\affiliation{Max-Planck-Institut für Astronomie, Königstuhl 17, D-69117 Heidelberg, Germany}
\email{rix@mpia.de}

\begin{abstract}
	\noindent
	Extremely metal-poor stars are intrinsically rare, but emerging methods exist to accurately classify them from all-sky Gaia XP low-resolution spectra.
	To assess their overall accuracy for targeting metal-poor stars, we present a high-resolution spectroscopic followup of 75 very metal-poor candidates selected from the catalog by R. Andrae, V. Chandra, and H. W. Rix.
	We discover 2 new extremely metal-poor ($\rm{[Fe / H]}<-3$) stars and 20 new very metal-poor ($\rm{[Fe/H]} < -2$) stars.
	Abundances of up to 22 elements are derived from 1D local thermodynamic equilibrium analysis and kinematic parameters are derived using Gaia astrometry and spectroscopic radial velocities.
	The chemodynamical properties are mostly consistent with expectations for halo stars, but we discover an Mg-enhanced CEMP star ($\mathrm{[Mg/Fe]} = 0.89$) and an Mg-poor star from an accreted ultra-faint dwarf galaxy.
	The Gaia XP metallicity estimates are consistent with our $\rm{[Fe/H]}$ measurements down to $\rm{[Fe/H]}\sim -3.0$, but estimates worsen in highly extincted regions.
	We find that 4 other XP-based metallicity catalogs succeed in mitigating contaminants and can also classify metal-poor stars robustly to $\rm{[Fe/H]}\sim -3.0$.
	Our results demonstrate the utility of Gaia XP spectra for identifying the most metal-poor stars across the Galaxy.
\end{abstract}


\section{Introduction} \label{sec:intro}
Studying the history of the Milky Way galaxy through its constituent stars relies on the observations of the most ancient, metal-poor stars in the galaxy \citep{eggenOldestDiskStars1974,freemanNewGalaxySignatures2002}. These metal-poor stars are assumed to be some of the oldest main-sequence and giant stars from the lack of heavy elements in their atmospheres and thus birth environments \citep{brommFirstStars2004, frebelNucleosyntheticSignaturesFirst2005}.
Through their unpolluted atmospheres and largely preserved orbits, they encode both the chemical signatures and dynamical history of their birth environment, providing a unique window into the evolution of Milky Way.

Metal-poor stars have been used to constrain the chemical evolution of the galaxy through probing the intertwined mechanisms of star formation, stellar yields, gas inflow, and feedback \citep[e.g.,][]{tinsleyStellarLifetimesAbundance1979, tinsleyEvolutionStarsGas1980, truranNewInterpretationHeavy1981, dekelOriginDwarfGalaxies1986, matteucciAbundanceRatiosEllipticals1994,mcwilliamAbundanceRatiosGalactic1997,  beersDiscoveryAnalysisVery2005}. Chemical abundances from large samples of metal-poor stars have provided observational constraints on the initial mass function, mixing processes, and explosion mechanisms of Population III stars \citep[e.g.,][]{,hegerNucleosynthesisEvolutionMassive2010,wanajoPhysicalConditionsRprocess2018,jonesNewModelElectroncapture2019,klessenFirstStarsFormation2023}.
More recent insights include larger demographics for observational constraints on \( r \)-process nucleosynthesis channels in the Milky Way galaxy and its satellites \citep{gudinRProcessAllianceChemodynamically2021a,kobayashiCanNeutronStar2023,frebelObservationsRProcessStars2023, guiglionObservationalConstraintsOrigin2024,lucchesiExtremelyMetalpoorStars2024,ouRiseRProcessGaiaSausage2024}, evidence for the intermediate process (\( i \)-process) channel of heavy element nucleosynthesis \citep{roedererDiverseOriginsNeutroncapture2016, clarksonPopIIIIprocess2018,hampelLearningIntermediateNeutroncapture2019,kochUnusualNeutroncaptureNucleosynthesis2019,skuladottirNeutroncaptureElementsDwarf2020, shejeelammalSpectroscopicStudyFour2022,goswamiPotentialCarbonEnhancedMetalpoor2024}, the early, proto-galactic dynamical history of the Milky Way  \citep{mashonkinaFormationMilkyWay2017,yuanDynamicalRelicsAncient2020, hortaEvidenceAPOGEEPresence2021, conroyBirthGalacticDisk2022, rixPoorOldHeart2022}, and the traces of unique, pre-Galactic carbon enrichment in the Milky Way and its satellites \citep{frebelNearFieldCosmologyExtremely2015,norrisMostMetalpoorStars2019,yoonIdentificationGroupIII2020,zepedaChemodynamicallyTaggedGroups2023,chitiEnrichmentExtragalacticFirst2024}.

Despite this, the discovery of metal-poor candidates has been historically difficult. Discovery of the most metal-poor stars \citep[e.g.,][]{norrisHE05574840UltraMetalPoor2007, caffauExtremelyPrimitiveStar2011, kellerSingleLowenergyIronpoor2014, melendez2MASSJ180820025104378Brightest2016, schlaufmanUltraMetalpoorStar2018} has been mainly driven by narrow-band photometric surveys \citep[e.g.,][]{schlaufmanBestBrightestMetalpoor2014,starkenburgPristineSurveyMining2017,dacostaSkyMapperDR11Search2019, galarzaJPLUSSearchingVery2022, placcoMiningSPLUSMetalpoor2022} or by serendipity across large spectroscopic surveys \citep[e.g.,][]{desilvaGALAHSurveyScientific2015,kollmeierSDSSVPioneeringPanoptic2017,placcoDECamMAGICSurvey2025}. Despite their success, these works together have an incomplete sky coverage and complex selection functions. A complete characterization of the most metal-poor stars in the Milky Way is vital to unravel its early chemodynamical history \citep{vennCouldUltraMetalPoorStars2008,frebelNearFieldCosmologyExtremely2015,klessenFirstStarsFormation2023, deasonGalacticArchaeologyGaia2024}.

With the release of \emph{Gaia} mission's Data Release 3 (DR3), contemporary studies have begun leveraging \emph{Gaia}'s low-resolution Blue Photometer and Red Photometer (BP/RP; XP) spectra \citep{angeliGaiaDataRelease2023} and Radial Velocity Spectrometer (RVS) spectra \citep{katzGaiaDataRelease2023} to perform preliminary, population-level analysis and candidate selection of the Galactic metal-poor population \citep{rixPoorOldHeart2022,zhangParameters220Million2023,luceyCarbonenhancedMetalpoorStar2023, xylakis-dornbuschMetallicitiesMore102024,yao200000Candidate2024,khalatyanTransferringSpectroscopicStellar2024,yangMetallicities20Million2025}. These studies heavily leverage \emph{Gaia}'s all-sky coverage with data-driven methods to classify metal-poor candidates through data-driven label transfer of high-confidence stellar labels and chemical abundances from large spectroscopic surveys such as APOGEE/SEGUE/SDSS \citep{yannySEGUESpectroscopicSurvey2009,majewskiApachePointObservatory2017a,kollmeierSloanDigitalSky2025,sdsscollaborationNineteenthDataRelease2025}, LEGUE/LAMOST \citep{cuiLargeSkyArea2012,dengLAMOSTExperimentGalactic2012}, and GALAH \citep{desilvaGALAHSurveyScientific2015,buderGALAHSurveyData2024}, or compiled observations of high-resolution metal-poor star samples \citep[e.g.,][]{sudaStellarAbundancesGalactic2008,sudaStellarAbundancesGalactic2014,sudaStellarAbundancesGalactic2017,abohalimaJINAbaseDatabaseChemical2018,yongHighresolutionSpectroscopicFollowup2021,liFourhundredVeryMetalpoor2022}.

One of the largest catalogs of metallicity estimations from \emph{Gaia} XP data products is presented by \citet{andraeRobustDatadrivenMetallicities2023} for over 175 million stars. They use the tree-based supervised machine learning \texttt{eXtreme Gradient Boosting} algorithm (\texttt{XGBoost}, \citealt{chenXGBoostScalableTree2016}) to translate stellar labels from SDSS/APOGEE and the high-resolution metal-poor sample presented in \citet{liFourhundredVeryMetalpoor2022} to the XP spectra of \emph{Gaia} DR3 sources.

It is important to test whether these approaches maintain a consistency in their estimations across the very-metal poor (VMP; \( \feh \le -2.0 \)) and extremely metal-poor (EMP; \( \feh \le  -3.0 \)) regimes, which are significantly more difficult to model accurately due to the high degeneracy with OBA-type stars and the low relative sample size of EMP/VMP stars. Determining the accuracy of these catalogs in this regime will support follow-up observations of the most metal-poor candidate stars, which is required to determine chemical abundances accurately. Chemical abundances of EMP stars and the lower metallicity regimes can then constrain our understanding of galactic chemical evolution in the early universe \citep{frebelNearFieldCosmologyExtremely2015,debennassutiLimitsPopulationIII2017,salvadoriProbingExistenceVery2019,kobayashiOriginElementsCarbon2020,klessenFirstStarsFormation2023}.

To aid in this effort, we have selected a sample of 75 metal-poor candidates from \citet{andraeRobustDatadrivenMetallicities2023}, obtained high-resolution spectroscopy from the European Southern Observatory (ESO), and we present a full chemodynamical analysis here.

This paper is organized as follows:
in Section \ref{sec:obs}, we explain our target selection, observations and data reduction process. Sections \ref{sec:params} explains our stellar parameter determinations.
Section \ref{sec:analysis} outlines our chemical abundance and kinematic analysis methods and results.
In Section \ref{sec:validate}, we compare our results to the \citet{andraeRobustDatadrivenMetallicities2023} catalog metallicities, assess the robustness and completeness of the catalog as compared with different works in the literature \citep{martinPristineSurveyXXIII2023,yao200000Candidate2024,khalatyanTransferringSpectroscopicStellar2024,yangMetallicities20Million2025}, and discuss the implications for targeting XP spectra.
Section \ref{sec:comment} provides specific comments on chemical abundance measurements.
Section \ref{sec:outliers} provides additional comments on detected carbon-enhanced metal-poor (CEMP), magnesium-unusual, and \( r \)-process enhanced (RPE) stars. Finally, we summarize our findings in Section \ref{sec:conclusion}.
Additionally, Appendix \ref{sec:darkcurrent} describes our issues with high dark current on the FEROS spectrograph.

\section{Observations}\label{sec:obs}
We selected candidate metal-poor red giant branch (RGB) stars from the vetted RGB catalog as described in \citet{andraeRobustDatadrivenMetallicities2023}. We additionally apply the following cuts to create a sample.

\begin{enumerate}
	\item \( G_{\mathrm{BP}} < 16 \) mag
	\item \( V < 13 \) mag
	\item \( \mh_{\mathrm{XP}} < -2.5 \) dex
	\item \( \log g_{\mathrm{XP}} < 3.0 \) dex
	\item \( 70 < \mathrm{RA} < 200 \) deg
	\item \( -70 < \mathrm{DEC} < 10  \) deg
\end{enumerate}
\begin{table}[t]
	\centering
	\begin{tabular}{lrr}
		\toprule
		Gaia DR3 Source ID  & \( \mathrm{[M  / H]}_{\mathrm{XG}} \) (dex) & \( b \) (deg) \\
		\midrule
		2928802421800482304 & $-2.64$                                     & $-8.066 $     \\
		2934166389275261824 & $-2.61$                                     & $-5.433 $     \\
		2935996457654663424 & $-2.60 $                                    & $ -3.466$     \\
		3113656611623601792 & $-2.57$                                     & $0.405  $     \\
		5242154747530026112 & $-2.65$                                     & $-1.522 $     \\
		5258578225752632320 & $-2.60 $                                    & $ -1.109$     \\
		5307768829791233536 & $-2.80 $                                    & $ -1.002$     \\
		5310532520992486912 & $-2.67$                                     & $-2.806 $     \\
		5333577528770381824 & $-2.61$                                     & $-1.747 $     \\
		5338100743508647808 & $-2.68$                                     & $-0.126 $     \\
		5350630404380049536 & $-2.51$                                     & $-0.646 $     \\
		5543467353546359936 & $-2.99$                                     & $1.744  $     \\
		5863544023161862144 & $-2.63$                                     & $1.518  $     \\
		5869608620084858240 & $-2.64$                                     & $1.896  $     \\
		6057362634399752192 & $-2.86$                                     & $-0.725 $     \\
		\bottomrule
	\end{tabular}
	\caption{Table of \emph{Gaia} DR3 source identifiers of all 15 OBA-type contaminants in our sample, along with their estimated metallicity \( \mh{}_{\mathrm{XG}} \) from \citet{andraeRobustDatadrivenMetallicities2023} and vertical position on the Galactic plane \( b \). All contaminants are concentrated along the Galactic plane along regions of high extinction.}\label{tab:bad}
\end{table}

This selection yielded 90 targets, all of which were observed during a single observing run during February and March 2024 using the FEROS instrument \citep[\( R = 48,000 \),][]{kauferFEROSFiberfedExtended1997} on the MPG/ESO 2.2m telescope at La Silla Observatory, Chile. The sky coordinate filters ensured that the targets were observable from La Silla Observatory in February, and the magnitude cuts ensured both valid metallicity estimates from \citet{andraeRobustDatadrivenMetallicities2023} and the targets were observable on the 2.2m telescope respectively. Of these selected targets, 22 stars have their abundances measured here for the first time.

The data were reduced automatically using the CERES reduction pipeline \citep{brahmCERESSetAutomated2017}, which performed bias and dark subtraction and wavelength calibration with lamp frames. We do not use the output radial velocities from CERES, as they occasionally fail to return consistent velocities with \emph{Gaia} RVS data if the magnitude is larger than 300 \( \mathrm{km\,s^{-1} } \).

Radial velocities are determined from our observations through the cross-correlation of 11 normalized spectral orders between 4000 and 6800 \AA{} against a high-S/N spectrum of HD 122563 through \texttt{PayneEchelle} as implemented in \texttt{LESSPayne}\footnote{\href{https://github.com/alexji/lesspayne}{github.com/alexji/LESSPayne}} \citep{tingPayneSelfconsistentInitio2019, jiLESSPayneLabelingEchelle2025}. We find that both \emph{Gaia} RVS data and \texttt{PayneEchelle} give similar heliocentric velocities to a median difference of \( 0.20\,\mathrm{km\,s^{-1} } \), well within the median uncertainty of \( 0.34\,\mathrm{km\,s^{-1} } \).

A preliminary analysis of our reduced high-resolution spectra revealed that 15 of the observed candidates were OBA-type contaminant stars.
The \emph{Gaia} identifiers for these stars are given in Table \ref{tab:bad}, alongside their reddening values and sky position.
All stars were in line with the galactic plane (\( |b| < 10 \)), with consequently high reddening values (\( E(B-V) > 0.2 \)). Their spectra either showed emission across \( \mathrm{H} \)-\( \alpha  \) or \( \mathrm{H} \)-\( \beta \), or wider hydrogen absorption features with minimal other extended features. These are consistent with OBA-type stellar atmospheres, and are expected to appear in our sample due to the degeneracy of reddened OBA stars with metal-poor stars \citep[see, e.g., Sections 3.7 \& 4.2 of][]{andraeRobustDatadrivenMetallicities2023}. We have removed these stars from our analyses.

\section{Stellar parameters}\label{sec:params}
\begin{figure}[t]
	\centering
	\includegraphics[width=\columnwidth]{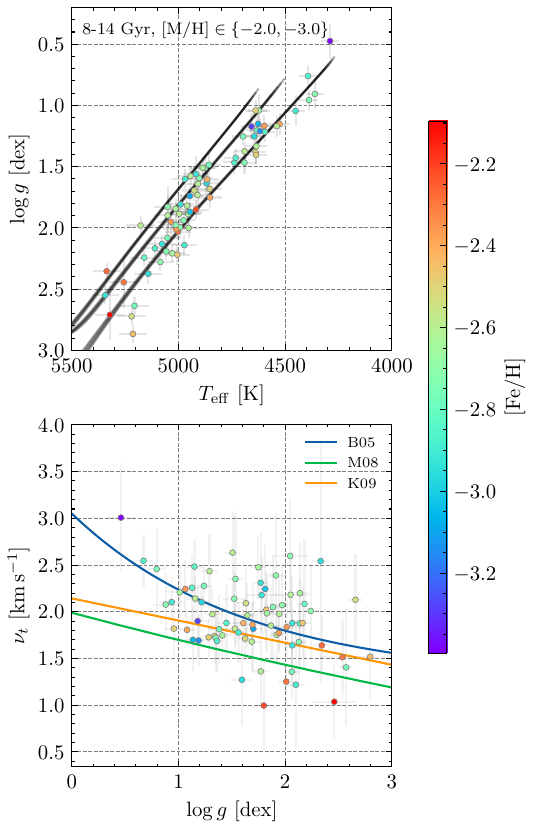}
	\caption{Stellar parameters for all analyzed compared to scaling relations. Top: effective temperature \( \teff \) versus surface gravity \( \log g \) compared to MIST isochrones \citep{choiMesaIsochronesStellar2016} of three different metallicities (\( \feh = -2.0, -2.5, -3.0 \)). We generally recover slightly higher or lower \( \teff \) and \( \logg \) values than expected from the isochrones. Bottom: surface gravity \( \log g \) vs microturbulence \( \nu_{t} \) compared to the empirical scaling relations given in \citet[][B05]{barklemHamburgESORprocess2005}, \citet[][M08]{marinoSpectroscopicPhotometricEvidence2008}, and \citet[][K09]{kirbyMultielementAbundanceMeasurements2009}.}
	\label{fig:params}
\end{figure}

Effective temperatures \( \teff \) were determined from the dereddened \emph{Gaia} \( BP-RP \) color index using the infrared flux method described in \citet{mucciarelliExploitingGaiaEDR32021} for red giants. The specific photometry was from \emph{Gaia} Data Release 3 (DR3) \citep{gaiacollaborationGaiaDataRelease2023}.
Extinction \( E(B-V) \) is determined from the dust skymap in \citet{schlegelMapsDustInfrared1998a} using corrections from \cite[][herein SFD]{schlaflyMeasuringReddeningSloan2011}, with conversions to the \emph{Gaia} photometry given in the auxiliary data of \emph{Gaia} DR3\footnote{\href{https://www.cosmos.esa.int/web/gaia/edr3-extinction-law}{cosmos.esa.int/web/gaia/edr3-extinction-law} (Baubsaiux C., ESA/Gaia/DPAC/CU5)}.
As a proxy for the dominant systematic uncertainties in the photometry and reddening, we added an uncertainty of 0.03 mag in quadrature with the statistical uncertainty on the corrected \( G_{\mathrm{BP}} -G_{\mathrm{RP}} \) color. Propagating uncertainties gives a typical \( \teff \) uncertainty of 50-65 K, which is similar to the scatter from other filter combinations.

Surface gravities were then determined from the logarithmic scaling relationship derived from the Stefan-Boltzmann law, given as:
\begin{eqnarray*}
	\log  g &=& \log g_{\odot} +\log (M / M_{\odot}) \\
	&&+ 4 \log (T_{\mathrm{eff}} / T_{\mathrm{eff},\odot})\\
	&&+ 0.4(M_{\mathrm{bol}} - M_{\mathrm{bol},\odot})
\end{eqnarray*}
where \( \log g \) is the surface gravity, \( M \) is the estimated mass of the star. For simplicity, we take the mass \( M \) to be \( 0.75 \pm 0.1\,\mathrm{M_{\odot}} \), as typical for an old, metal-poor red giant. We take the Solar stellar parameters as \( T_{\mathrm{eff, \odot}} = 5770\,\mathrm{K}, \log g_{\odot} = 4.44, M_{\mathrm{bol,\odot}} =4.76 \). Bolometric magnitude corrections are taken from the grid for RGB stars described in \citet{casagrandeUseGaiaMagnitudes2018}, with the absolute \( G \) magnitude being used as the preceding magnitude. Geometric distances from \citet{bailer-jonesEstimatingDistancesParallaxes2021} are used instead of inverse parallax for determining distance, and are the dominant source of uncertainty. The typical \( \logg \) uncertainty is \( 0.07 \)-\( 0.08 \) dex, given by propagating through this equation.

Using the SFD dustmap led to spurious dereddenings and thus erroneous stellar parameters for 11 stars. After deriving the metallicity of these stars, the previously derived \( \teff \) and \( \logg \) appeared physically inconsistent with the MIST isochrones \citep{choiMesaIsochronesStellar2016} (i.e. hot, low \( \log g \) metal-rich stars). We found these stars had particularly high reddening values for their short distances driven by their sky position along the Galactic plane. Since the SFD dustmap assumes they are behind the dust layer on the 2D dustmap, we use the three-dimensional Milky Way dustmaps from \citet{edenhoferParsecscaleGalactic3D2024} and \citet{green3DDustMap2019} to more accurately determine \( E(B-V) \) by factoring their proximity into the derivation. We selected these stars based on the criteria
\begin{eqnarray*}
	\bigl[(E(B-V) \ge 0.2) \,\mid \, (-15 < l < 15)\bigr] \wedge (d_{\mathrm{kpc}} < 1.25)
\end{eqnarray*}

which yields 11 stars, 5 of which are covered by the 3D dustmaps.
For the 6 additional stars which are outside of both dustmaps' coverage, we instead calculate the surface gravity by fitting to the red giant branch of a 12 Gyr MIST isochrone track \citep{choiMesaIsochronesStellar2016}. We interpolate for \( \logg \) based on the color and absolute \emph{Gaia} \( G \) magnitude \( M_{G} \). The isochrone metallicity is chosen based on how well isochrones ranging from \( \mathrm{[M / H]} \in [-4.0,-3.5,-3.0,-2.5,-2.0] \) predict \( M_{G} \). The differences between the \( \teff \) from the best fitting metallicity isochrone and the next two best fitting isochrones are additionally added in quadrature to the \( \teff \) uncertainty. Since \( \logg \) in these stars now heavily depends on accurate dereddening, we add an additional 0.1 dex systematic uncertainty in quadrature to the \( \logg \). The typical \( \logg \) uncertainty for these stars is \( 0.15 \)-\( 0.20 \) dex.

Metallicity \( \feh\) (which is taken as \( \mh \)) is then determined simultaneously with the microturbulence \( \nu _t \) from the equivalent widths of Fe II lines, where we hold the constraint that the model metallicity \( \mh  \) is equal to the average abundance from the Fe I lines.

\begin{longrotatetable}
	\begin{deluxetable}{lccccccccccrr}
		\tablecolumns{13}
		\tabletypesize{\scriptsize}
		\tablecaption{\label{tab:params}Stellar Parameters}
		\tablehead{Gaia DR3 ID  & \( G_{0} \) (mag) & \( BP_{0} \) (mag) & \( RP_{0} \) (mag) & \( E(B-V) \)  & \( M_{G} \) (mag) & \( \teff \) (K) & \( \logg \) (dex)     & \( \vt \) \( (\mathrm{km\,s^{-1}} ) \) & \( \mathrm{[M / H]} \) (dex) & \( v_{\mathrm{rad}} \) \( (\mathrm{km\,s^{-1}} ) \) & Dustmap   & New}
		\startdata
		582484053094720640  & 12.71             & 13.13              & 12.12              & 0.03               & $0.58  \pm 0.11$  & $5142 \pm 64$   & $2.33 \pm 0.08$ & $2.54 \pm 1.30$                                    & $-2.95 \pm 0.29$             & $94.00\pm 0.08$                                     & SFD       & Y   \\
		2895944715088173184 & 12.34             & 12.75              & 11.74              & 0.04               & $-0.43 \pm 0.08$  & $5177 \pm 67$   & $1.92 \pm 0.07$ & $1.75 \pm 0.32$                                    & $-2.60 \pm 0.33$             & $112.6\pm 0.06$                                     & SFD       & N   \\
		2909738294618626048 & 12.33             & 12.90              & 11.61              & 0.03               & $-1.98 \pm 0.16$  & $4598 \pm 49$   & $1.07 \pm 0.09$ & $2.24 \pm 0.31$                                    & $-2.36 \pm 0.51$             & $190.8\pm 0.04$                                     & SFD       & N   \\
		2938430256702303360 & 11.89             & 12.48              & 11.14              & 0.12               & $-1.04 \pm 0.09$  & $4691 \pm 51$   & $1.32 \pm 0.08$ & $1.97 \pm 0.27$                                    & $-2.59 \pm 0.30$             & $-94.1\pm 0.03$                                     & SFD       & N   \\
		2960365655412742784 & 11.47             & 12.02              & 10.76              & 0.03               & $-1.49 \pm 0.11$  & $4634 \pm 50$   & $1.29 \pm 0.08$ & $2.43 \pm 0.40$                                    & $-2.61 \pm 0.28$             & $182.7\pm 0.03$                                     & SFD       & N   \\
		2970270979572570240 & 11.05             & 11.49              & 10.43              & 0.05               & $0.08  \pm 0.06$  & $5104 \pm 63$   & $2.07 \pm 0.07$ & $1.88 \pm 0.49$                                    & $-2.93 \pm 0.18$             & $320.4\pm 0.03$                                     & SFD       & N   \\
		3010477130501851136 & 12.26             & 12.74              & 11.60              & 0.12               & $0.44  \pm 0.07$  & $5100 \pm 63$   & $2.07 \pm 0.07$ & $1.64 \pm 0.38$                                    & $-2.93 \pm 0.26$             & $53.70\pm 0.05$                                     & Bayestar  & Y   \\
		3319092078574550400 & 11.81             & 12.45              & 11.02              & 0.16               & $-0.23 \pm 0.08$  & $4623 \pm 50$   & $1.53 \pm 0.07$ & $1.82 \pm 0.27$                                    & $-2.59 \pm 0.27$             & $209.1\pm 0.03$                                     & Bayestar  & Y   \\
		3474451177494568320 & 12.30             & 12.79              & 11.63              & 0.06               & $-1.00 \pm 0.12$  & $4904 \pm 58$   & $1.53 \pm 0.08$ & $2.35 \pm 0.71$                                    & $-2.73 \pm 0.34$             & $-85.8\pm 0.05$                                     & SFD       & N   \\
		3480807557294289664 & 12.42             & 12.90              & 11.77              & 0.06               & $-0.36 \pm 0.11$  & $4954 \pm 59$   & $1.82 \pm 0.08$ & $2.02 \pm 0.42$                                    & $-2.46 \pm 0.27$             & $7.20 \pm 0.04$                                     & SFD       & Y   \\
		3492757569356904320 & 10.50             & 10.99              & 9.83               & 0.05               & $-0.83 \pm 0.06$  & $4883 \pm 57$   & $1.63 \pm 0.07$ & $1.71 \pm 0.19$                                    & $-2.51 \pm 0.23$             & $349.5\pm 0.01$                                     & SFD       & N   \\
		3504642358045116416 & 12.43             & 12.96              & 11.73              & 0.12               & $-0.08 \pm 0.09$  & $4918 \pm 58$   & $1.80 \pm 0.08$ & $1.00 \pm 0.49$                                    & $-2.21 \pm 0.37$             & $31.10\pm 0.04$                                     & SFD       & N   \\
		3513718345495642368 & 12.06             & 12.52              & 11.41              & 0.05               & $-0.54 \pm 0.12$  & $4989 \pm 60$   & $1.77 \pm 0.08$ & $2.31 \pm 0.55$                                    & $-2.93 \pm 0.29$             & $308.3\pm 0.04$                                     & SFD       & N   \\
		3520836313191203584 & 11.81             & 12.29              & 11.15              & 0.08               & $0.16  \pm 0.09$  & $5002 \pm 60$   & $2.01 \pm 0.08$ & $1.25 \pm 0.24$                                    & $-2.29 \pm 0.31$             & $-79.7\pm 0.03$                                     & SFD       & N   \\
		3521175856125903616 & 11.51             & 12.06              & 10.80              & 0.04               & $-1.74 \pm 0.22$  & $4658 \pm 51$   & $1.18 \pm 0.11$ & $1.90 \pm 0.39$                                    & $-3.29 \pm 0.25$             & $332.8\pm 0.03$                                     & SFD       & N   \\
		3523787024442853504 & 12.50             & 13.06              & 11.79              & 0.04               & $-1.33 \pm 0.18$  & $4639 \pm 49$   & $1.34 \pm 0.10$ & $1.74 \pm 0.30$                                    & $-2.51 \pm 0.31$             & $64.40\pm 0.03$                                     & SFD       & Y   \\
		3549787824444153856 & 12.51             & 12.98              & 11.85              & 0.06               & $-0.50 \pm 0.12$  & $4961 \pm 59$   & $1.77 \pm 0.08$ & $1.36 \pm 0.37$                                    & $-2.61 \pm 0.34$             & $291.5\pm 0.05$                                     & SFD       & N   \\
		3552621991760308864 & 12.59             & 13.07              & 11.93              & 0.03               & $-0.97 \pm 0.18$  & $4869 \pm 57$   & $1.59 \pm 0.10$ & $1.27 \pm 0.49$                                    & $-2.97 \pm 0.37$             & $19.30\pm 0.04$                                     & SFD       & N   \\
		3594461329774715392 & 12.51             & 13.17              & 11.73              & 0.03               & $-2.12 \pm 0.25$  & $4361 \pm 43$   & $0.88 \pm 0.12$ & $2.08 \pm 0.34$                                    & $-2.72 \pm 0.33$             & $387.3\pm 0.03$                                     & SFD       & N   \\
		3598383734427539840 & 12.74             & 13.20              & 12.10              & 0.04               & $0.21  \pm 0.12$  & $4972 \pm 59$   & $2.10 \pm 0.08$ & $1.22 \pm 0.69$                                    & $-2.89 \pm 0.38$             & $91.30\pm 0.09$                                     & SFD       & Y   \\
		3600053544696445568 & 11.82             & 12.30              & 11.17              & 0.02               & $-1.00 \pm 0.15$  & $4867 \pm 57$   & $1.61 \pm 0.09$ & $1.88 \pm 0.21$                                    & $-2.42 \pm 0.35$             & $-46.0\pm 0.02$                                     & SFD       & N   \\
		3621673727165280384 & 11.92             & 12.35              & 11.30              & 0.05               & $0.22  \pm 0.06$  & $5111 \pm 64$   & $2.14 \pm 0.07$ & $1.88 \pm 0.49$                                    & $-2.86 \pm 0.27$             & $166.5\pm 0.04$                                     & SFD       & N   \\
		3625358358132976128 & 12.13             & 12.61              & 11.48              & 0.04               & $-0.74 \pm 0.12$  & $4912 \pm 58$   & $1.69 \pm 0.08$ & $1.68 \pm 0.30$                                    & $-2.61 \pm 0.27$             & $163.0\pm 0.04$                                     & SFD       & N   \\
		3636133400286350976 & 11.95             & 12.44              & 11.30              & 0.03               & $-1.16 \pm 0.21$  & $4876 \pm 57$   & $1.52 \pm 0.11$ & $2.14 \pm 0.30$                                    & $-2.72 \pm 0.21$             & $-39.3\pm 0.03$                                     & SFD       & N   \\
		3684987592422300800 & 12.60             & 13.02              & 11.99              & 0.03               & $0.39  \pm 0.09$  & $5085 \pm 63$   & $2.24 \pm 0.08$ & $2.01 \pm 0.51$                                    & $-2.71 \pm 0.35$             & $125.2\pm 0.05$                                     & SFD       & N   \\
		3700301556014893952 & 12.72             & 13.16              & 12.10              & 0.02               & $0.23  \pm 0.13$  & $5007 \pm 61$   & $2.16 \pm 0.08$ & $1.88 \pm 0.39$                                    & $-2.47 \pm 0.31$             & $100.6\pm 0.05$                                     & SFD       & N   \\
		3771112024693016832 & 12.37             & 12.84              & 11.72              & 0.05               & $-0.40 \pm 0.11$  & $4954 \pm 59$   & $1.83 \pm 0.08$ & $1.98 \pm 0.42$                                    & $-2.66 \pm 0.34$             & $414.0\pm 0.05$                                     & SFD       & N   \\
		3800139440903714688 & 11.91             & 12.35              & 11.28              & 0.03               & $-0.46 \pm 0.09$  & $4997 \pm 61$   & $1.87 \pm 0.08$ & $1.85 \pm 0.38$                                    & $-2.60 \pm 0.23$             & $-143.\pm 0.03$                                     & SFD       & N   \\
		3818459160048340352 & 12.06             & 12.61              & 11.34              & 0.03               & $-1.88 \pm 0.20$  & $4627 \pm 50$   & $1.13 \pm 0.11$ & $1.70 \pm 0.35$                                    & $-3.06 \pm 0.28$             & $-98.5\pm 0.03$                                     & SFD       & N   \\
		3822507802379684608 & 12.30             & 12.76              & 11.65              & 0.05               & $-0.26 \pm 0.11$  & $4976 \pm 60$   & $1.89 \pm 0.08$ & $2.05 \pm 0.56$                                    & $-2.73 \pm 0.28$             & $234.4\pm 0.04$                                     & SFD       & Y   \\
		3864303022490267136 & 11.46             & 11.90              & 10.85              & 0.03               & $-0.02 \pm 0.07$  & $5053 \pm 62$   & $2.06 \pm 0.07$ & $1.36 \pm 0.28$                                    & $-2.77 \pm 0.28$             & $67.20\pm 0.03$                                     & SFD       & N   \\
		3874150420427978624 & 11.81             & 12.28              & 11.16              & 0.04               & $-0.87 \pm 0.17$  & $4910 \pm 58$   & $1.65 \pm 0.10$ & $1.96 \pm 0.35$                                    & $-2.61 \pm 0.23$             & $85.20\pm 0.02$                                     & SFD       & Y   \\
		3874671039183974272 & 12.09             & 12.64              & 11.37              & 0.02               & $-2.20 \pm 0.28$  & $4624 \pm 50$   & $1.01 \pm 0.13$ & $2.21 \pm 0.29$                                    & $-2.67 \pm 0.25$             & $-65.8\pm 0.02$                                     & SFD       & Y   \\
		3892663864073428480 & 12.69             & 13.21              & 12.00              & 0.02               & $-1.24 \pm 0.16$  & $4738 \pm 53$   & $1.45 \pm 0.09$ & $1.88 \pm 0.41$                                    & $-2.86 \pm 0.29$             & $60.60\pm 0.05$                                     & SFD       & Y   \\
		3910233647567741056 & 11.94             & 12.38              & 11.32              & 0.02               & $-0.22 \pm 0.09$  & $4994 \pm 60$   & $1.97 \pm 0.08$ & $2.07 \pm 0.43$                                    & $-2.70 \pm 0.29$             & $-215.\pm 0.03$                                     & SFD       & N   \\
		4757476003231981952 & 11.75             & 12.31              & 11.03              & 0.04               & $-2.25 \pm 0.15$  & $4638 \pm 50$   & $0.96 \pm 0.09$ & $1.82 \pm 0.31$                                    & $-2.50 \pm 0.40$             & $40.40\pm 0.03$                                     & SFD       & N   \\
		4761418199093214592 & 12.38             & 12.91              & 11.68              & 0.02               & $-1.44 \pm 0.15$  & $4691 \pm 52$   & $1.35 \pm 0.09$ & $1.72 \pm 0.36$                                    & $-2.75 \pm 0.28$             & $147.0\pm 0.04$                                     & SFD       & N   \\
		4762819324800428928 & 11.20             & 11.72              & 10.52              & 0.02               & $-1.46 \pm 0.08$  & $4732 \pm 52$   & $1.36 \pm 0.07$ & $1.69 \pm 0.33$                                    & $-2.86 \pm 0.31$             & $323.9\pm 0.02$                                     & SFD       & N   \\
		4769570635432428032 & 12.64             & 13.09              & 12.01              & 0.04               & $0.25  \pm 0.06$  & $5030 \pm 61$   & $2.14 \pm 0.07$ & $2.20 \pm 0.55$                                    & $-2.64 \pm 0.31$             & $128.5\pm 0.05$                                     & SFD       & N   \\
		4800721449115828352 & 11.33             & 11.90              & 10.62              & 0.02               & $-1.83 \pm 0.12$  & $4596 \pm 49$   & $1.15 \pm 0.08$ & $2.48 \pm 0.29$                                    & $-2.84 \pm 0.19$             & $231.7\pm 0.02$                                     & SFD       & N   \\
		4801358478664227456 & 12.68             & 13.08              & 12.09              & 0.04               & $1.16  \pm 0.04$  & $5207 \pm 67$   & $2.57 \pm 0.07$ & $1.40 \pm 0.63$                                    & $-2.78 \pm 0.24$             & $488.5\pm 0.04$                                     & SFD       & N   \\
		4811990171989821824 & 11.74             & 12.32              & 11.01              & 0.01               & $-1.97 \pm 0.13$  & $4526 \pm 47$   & $1.08 \pm 0.09$ & $1.81 \pm 0.17$                                    & $-2.39 \pm 0.27$             & $-21.1\pm 0.02$                                     & SFD       & N   \\
		5223645951042953984 & 12.55             & 12.97              & 11.95              & 0.08               & $0.85  \pm 0.04$  & $5257 \pm 69$   & $2.54 \pm 0.21$ & $1.51 \pm 0.41$                                    & $-2.29 \pm 0.35$             & $306.6\pm 0.04$                                     & SFD+Iso & Y   \\
		5232963831059764224 & 11.50             & 12.11              & 10.73              & 0.24               & $-0.96 \pm 0.06$  & $4919 \pm 58$   & $1.78 \pm 0.15$ & $2.18 \pm 0.41$                                    & $-2.93 \pm 0.26$             & $36.00\pm 0.03$                                     & SFD+Iso & Y   \\
		5281352715015701248 & 12.20             & 12.90              & 11.39              & 0.10               & $-2.06 \pm 0.13$  & $4389 \pm 44$   & $0.80 \pm 0.08$ & $2.46 \pm 0.44$                                    & $-2.76 \pm 0.34$             & $174.2\pm 0.03$                                     & SFD       & N   \\
		5283390938997970176 & 12.10             & 12.67              & 11.38              & 0.05               & $-1.39 \pm 0.13$  & $4636 \pm 50$   & $1.28 \pm 0.09$ & $1.73 \pm 0.24$                                    & $-2.49 \pm 0.30$             & $61.30\pm 0.03$                                     & SFD       & N   \\
		5296305287178801280 & 11.58             & 12.01              & 10.96              & 0.08               & $0.83  \pm 0.03$  & $5187 \pm 65$   & $2.34 \pm 0.07$ & $1.64 \pm 0.31$                                    & $-2.27 \pm 0.24$             & $-7.30\pm 0.02$                                     & Edenhofer & N   \\
		5346852929097845632 & 12.00             & 12.44              & 11.38              & 0.09               & $2.00  \pm 0.06$  & $5210 \pm 67$   & $2.80 \pm 0.07$ & $1.52 \pm 0.46$                                    & $-2.44 \pm 0.30$             & $310.7\pm 0.03$                                     & Edenhofer & Y   \\
		5349032191191937920 & 12.51             & 13.00              & 11.84              & 0.21               & $0.67  \pm 0.05$  & $5322 \pm 70$   & $2.46 \pm 0.21$ & $1.04 \pm 0.40$                                    & $-2.10 \pm 0.32$             & $50.80\pm 0.04$                                     & SFD+Iso & Y   \\
		5372831262118796288 & 12.71             & 13.21              & 12.04              & 0.13               & $-0.09 \pm 0.09$  & $5052 \pm 62$   & $2.05 \pm 0.17$ & $2.18 \pm 0.41$                                    & $-2.62 \pm 0.30$             & $211.5\pm 0.05$                                     & SFD+Iso & Y   \\
		5444147613512194176 & 11.14             & 11.86              & 10.31              & 0.08               & $-2.86 \pm 0.30$  & $4290 \pm 41$   & $0.46 \pm 0.14$ & $3.01 \pm 0.58$                                    & $-3.40 \pm 0.24$             & $58.00\pm 0.02$                                     & SFD       & Y   \\
		5449533055822718848 & 11.75             & 12.36              & 11.00              & 0.06               & $-1.56 \pm 0.13$  & $4539 \pm 48$   & $1.16 \pm 0.09$ & $2.14 \pm 0.23$                                    & $-2.59 \pm 0.24$             & $245.5\pm 0.02$                                     & SFD       & N   \\
		5487737374038584704 & 11.65             & 12.14              & 10.98              & 0.16               & $0.84  \pm 0.03$  & $5161 \pm 65$   & $2.18 \pm 0.07$ & $2.08 \pm 0.69$                                    & $-2.81 \pm 0.25$             & $174.1\pm 0.04$                                     & SFD       & N   \\
		5489119425795568256 & 12.69             & 13.21              & 11.99              & 0.15               & $-0.14 \pm 0.10$  & $5013 \pm 61$   & $1.75 \pm 0.08$ & $2.48 \pm 0.33$                                    & $-2.64 \pm 0.34$             & $147.7\pm 0.05$                                     & SFD       & N   \\
		5491893184393739776 & 11.78             & 12.32              & 11.08              & 0.14               & $-0.74 \pm 0.05$  & $4944 \pm 58$   & $1.51 \pm 0.07$ & $2.63 \pm 0.45$                                    & $-2.60 \pm 0.25$             & $229.6\pm 0.03$                                     & SFD       & N   \\
		5502084317152803200 & 10.89             & 11.35              & 10.25              & 0.05               & $-0.16 \pm 0.04$  & $5011 \pm 61$   & $1.94 \pm 0.07$ & $1.77 \pm 0.22$                                    & $-2.36 \pm 0.23$             & $8.90 \pm 0.02$                                     & SFD       & N   \\
		5508686643961355648 & 11.21             & 11.81              & 10.45              & 0.10               & $-1.52 \pm 0.10$  & $4633 \pm 50$   & $1.13 \pm 0.08$ & $2.26 \pm 0.35$                                    & $-2.83 \pm 0.24$             & $352.3\pm 0.02$                                     & SFD       & N   \\
		5518045893097566976 & 11.93             & 12.48              & 11.21              & 0.20               & $-0.39 \pm 0.06$  & $5035 \pm 61$   & $2.01 \pm 0.17$ & $1.84 \pm 0.30$                                    & $-2.37 \pm 0.31$             & $84.80\pm 0.03$                                     & SFD+Iso & N   \\
		5553564207478749568 & 11.20             & 11.86              & 10.42              & 0.05               & $-2.60 \pm 0.14$  & $4393 \pm 44$   & $0.67 \pm 0.09$ & $2.55 \pm 0.28$                                    & $-2.86 \pm 0.26$             & $65.80\pm 0.02$                                     & SFD       & Y   \\
		5633706540578910336 & 12.17             & 12.66              & 11.51              & 0.10               & $0.01  \pm 0.07$  & $5014 \pm 60$   & $1.91 \pm 0.07$ & $2.39 \pm 0.66$                                    & $-2.71 \pm 0.30$             & $50.40\pm 0.04$                                     & SFD       & N   \\
		5652695827945108864 & 10.62             & 11.23              & 9.87               & 0.10               & $-1.25 \pm 0.06$  & $4622 \pm 50$   & $1.24 \pm 0.07$ & $2.28 \pm 0.23$                                    & $-2.76 \pm 0.20$             & $390.7\pm 0.01$                                     & Bayestar  & Y   \\
		5660010981883926912 & 12.73             & 13.21              & 12.08              & 0.06               & $-0.06 \pm 0.11$  & $4952 \pm 59$   & $1.94 \pm 0.08$ & $1.98 \pm 0.40$                                    & $-2.62 \pm 0.29$             & $423.2\pm 0.05$                                     & SFD       & Y   \\
		5731383034718059520 & 11.78             & 12.35              & 11.05              & 0.05               & $-1.62 \pm 0.19$  & $4618 \pm 50$   & $1.19 \pm 0.10$ & $1.69 \pm 0.27$                                    & $-3.15 \pm 0.29$             & $198.6\pm 0.02$                                     & SFD       & N   \\
		5735381717990313216 & 11.59             & 12.07              & 10.93              & 0.06               & $-0.64 \pm 0.11$  & $4947 \pm 59$   & $1.70 \pm 0.08$ & $1.82 \pm 0.47$                                    & $-3.10 \pm 0.24$             & $202.1\pm 0.03$                                     & SFD       & N   \\
		5739501549403972736 & 12.13             & 12.52              & 11.55              & 0.04               & $1.33  \pm 0.04$  & $5253 \pm 68$   & $2.66 \pm 0.07$ & $2.13 \pm 0.49$                                    & $-2.54 \pm 0.28$             & $155.7\pm 0.03$                                     & SFD       & Y   \\
		5755996908175855232 & 9.98              & 10.53              & 9.27               & 0.03               & $-1.69 \pm 0.09$  & $4644 \pm 51$   & $1.22 \pm 0.08$ & $2.10 \pm 0.20$                                    & $-2.97 \pm 0.19$             & $169.5\pm 0.01$                                     & SFD       & N   \\
		6059579009301748352 & 10.43             & 10.99              & 9.71               & 0.17               & $0.23  \pm 0.03$  & $4934 \pm 60$   & $1.81 \pm 0.07$ & $2.24 \pm 0.47$                                    & $-3.03 \pm 0.18$             & $240.3\pm 0.02$                                     & Edenhofer & Y   \\
		6076898344653529344 & 12.02             & 12.55              & 11.32              & 0.18               & $0.01  \pm 0.06$  & $5050 \pm 62$   & $2.05 \pm 0.17$ & $2.60 \pm 0.61$                                    & $-2.72 \pm 0.28$             & $193.7\pm 0.02$                                     & SFD+Iso & N   \\
		6133678941862384768 & 12.45             & 13.10              & 11.66              & 0.08               & $-1.90 \pm 0.25$  & $4451 \pm 45$   & $0.94 \pm 0.12$ & $2.10 \pm 0.29$                                    & $-2.90 \pm 0.31$             & $243.2\pm 0.03$                                     & SFD       & N   \\
		6134455579331553536 & 12.51             & 13.05              & 11.82              & 0.10               & $-0.37 \pm 0.13$  & $4853 \pm 56$   & $1.69 \pm 0.09$ & $1.86 \pm 0.23$                                    & $-2.40 \pm 0.21$             & $-12.5\pm 0.03$                                     & SFD       & N   \\
		6139117847928502528 & 12.62             & 13.12              & 11.94              & 0.08               & $-0.67 \pm 0.13$  & $4927 \pm 59$   & $1.63 \pm 0.09$ & $2.09 \pm 0.37$                                    & $-2.54 \pm 0.30$             & $187.6\pm 0.03$                                     & SFD       & N   \\
		6146588965016890752 & 12.20             & 12.72              & 11.51              & 0.10               & $-1.12 \pm 0.15$  & $4886 \pm 57$   & $1.41 \pm 0.09$ & $1.75 \pm 0.25$                                    & $-2.64 \pm 0.26$             & $114.2\pm 0.05$                                     & SFD       & N   \\
		6150224495918526848 & 12.17             & 12.63              & 11.54              & 0.07               & $0.37  \pm 0.07$  & $5056 \pm 61$   & $2.13 \pm 0.07$ & $1.68 \pm 0.42$                                    & $-2.74 \pm 0.25$             & $343.2\pm 0.05$                                     & SFD       & N   \\
		6158271340483740288 & 12.39             & 12.90              & 11.72              & 0.06               & $-1.35 \pm 0.21$  & $4857 \pm 57$   & $1.38 \pm 0.11$ & $1.81 \pm 0.38$                                    & $-2.78 \pm 0.34$             & $37.90\pm 0.05$                                     & SFD       & N   \\
		6167441439259305088 & 12.10             & 12.57              & 11.45              & 0.05               & $-1.04 \pm 0.12$  & $4969 \pm 60$   & $1.56 \pm 0.08$ & $1.78 \pm 0.33$                                    & $-2.89 \pm 0.24$             & $142.7\pm 0.04$                                     & SFD       & Y   \\
		\enddata
		\tablecomments{``Dustmap" refers to the specific dustmap used for \( E(B-V) \) estimation. SFD = \citet{schlaflyMeasuringReddeningSloan2011} (+Iso means isochrone used for \( \logg \)), Bayestar = \citet{greenGalacticReddening3D2018}, Edenhofer = \citet{edenhoferParsecscaleGalactic3D2024}.}
		\vspace{-3em}
	\end{deluxetable}
\end{longrotatetable}

Microturbulence \( \nu_{\mathrm{t}} \) is determined from the standard condition of independence of Fe II abundance \citep{jiSouthernStellarStream2020} with respect to reduced equivalent width of the modelled absorption feature \citep{nissenHighprecisionStellarAbundances2018}. In other words, we set the microturbulence by requiring both strong and weak atomic transitions to predict the same iron chemical abundance. The uncertainty is estimated by varying \( \nu_{t} \) until the slope changes by one standard error on the slope. The typical \( \nu_{\mathrm{t}} \) uncertainty is \( 0.39\,\mathrm{km\,s^{-1} } \), but is greater than \( 0.6\,\mathrm{km\,s^{-1} } \) in 7 low S/N stars with few (\( \sim 10 \)) recovered Fe II lines.

All of our measured stellar parameters are shown in Figure \ref{fig:params} and Table \ref{tab:params}. The top panel shows \( \teff \) versus \( \logg \) for our stars, which match well to the MIST isochrones. The bottom panel shows \( \nu_{\mathrm{t}} \) versus \( \logg \) for our stars compared to three empirical \( \logg \) vs. \( \nu_{\mathrm{t}} \) fits, which resemble the empirical fit to cool metal-poor giants from \citet{barklemHamburgESORprocess2005}.

\section{Analysis} \label{sec:analysis}
\subsection{Abundance analysis}

To derive chemical abundances, we perform a standard abundance analysis with the 2017 version of the 1D local thermodynamic equilibrium (LTE) radiative transfer code \texttt{MOOG}\footnote{\href{https://github.com/alexji/moog17scat}{github.com/alexji/moog17scat}} \citep{snedenCarbonNitrogenAbundances1973, sobeckAbundancesNeutroncaptureSpecies2011} and the \texttt{ATLAS} model atmospheres \citep{castelliNewGridsATLAS92003}.

We perform equivalent width measurements for 10 elements by fitting Gaussian line profiles to the absorption lines in the spectra, and perform synthetic spectra fits for the remaining 11 elements with weaker or blended absorption features, namely scandium, vanadium, cobalt, strontium, yttrium, zirconium, barium, lanthanum, and europium. For our reported abundances, we choose to use Fe I for neutral species and Fe II for ionized species to derive \( \mathrm{[X / Fe]} \) values.

As standard for halo stars in the Milky Way, we assumed an \( \alpha  \)-process elemental abundance of \( \afe = +0.4 \). Both methods were utilized via the tools \texttt{SMHR}\footnote{\href{https://github.com/andycasey/smhr}{github.com/andycasey/smhr}} \citep{caseyTaleTidalTales2014,caseySmhrAutomaticCurveofgrowth2025} and the \texttt{autosmh} component in \texttt{LESSPayne}.

The base linelist used for our spectroscopic analysis is described in Section 4.1 of \citet{jiSouthernStellarStream2020}, except we exclude features above 6800\,\AA{} due to the lower wavelength range of the FEROS spectrograph. While we include features below 4000\,\AA{} these are generally not recovered due to the poor S/N in the bluer spectral orders. We additionally exclude the dysprosium features and the additional europium features at 4352\,\AA{} and 4452\,\AA{} lines, as these weak features remain difficult to detect at the metallicities of our sample, which is \( \sim\!1 \) dex lower than the stars studied in \citet{jiSouthernStellarStream2020}. Limitations of specific chemical element measurements are described in Section \ref{sec:comment}.

Upper limits are derived with spectral synthesis if a \( 3 \sigma  \) detection is not met. For each feature, a synthetic spectrum was fit to match the continuum, radial velocity, and smoothing of the observed spectrum. Holding continuum and smoothing fixed, the abundance was increased until \( \Delta  \chi ^2 = 25 \), which is formally a \( 5 \sigma  \) upper limit. While this does not provide an uncertainty for the continuum placement, we do not return any upper limits for the molecular feature \( \mathrm{CH} \) where this would be significant.

Chemical abundances errors are determined line-by-line for the propagated stellar parameters error, the line-to-line scatter, and statistical uncertainties \citep{jiNearlyPristineStar2025}.
Abundances \( \mathrm{[X / H]} \) are determined from the weighted mean of each line measurement, such that \( \log \epsilon_{X} \equiv \hat{x} = \Sigma_{i} w_{i} x_{i} / \Sigma_{i} w_{i} \).
These weights are defined per line, such that \( w_{i} = (\sigma_{\mathrm{stat},i}^2 + \sigma_{\mathrm{SP},i}^2+  s_{X}^2)^{-1} \), for statistical error \( \sigma_{\mathrm{stat},i} \) and systematic error \( s_{\mathrm{X}} \). The stellar parameter uncertainty is propagated, where \( \sigma_{\mathrm{SP},i}^2 = \Sigma_{k} (\Delta_{k}X)_{i}\) for each abundance change \( \Delta_{k} X \) given by the uncertainty of each stellar parameter \( k \).
We estimate the systematic error term \( s_{\mathrm{X}} \) from the line-to-line scatter excluding the stellar parameter uncertainties, where individual line measurements \( x_{i} \) are modelled as \(x_{i}\sim  \mathcal{N}(x_{\mathrm{True}}, \sigma_{\mathrm{stat},i}^2 + s_{\mathrm{X}}^2)\) for true abundance \( x_{\mathrm{True}} \).
To obtain errors on \( \log \epsilon \) and thereby \( \mathrm{[X / H]} \), we combine the statistical error, systematic error, and the weighted mean of abundance differences from each stellar parameter uncertainty, such that \( e(\mathrm{[X / H]})^2 = 1/\Sigma_{i}\overline{w}_{i}  +  \Sigma_{k} [(\Sigma_{i} w_{i} \Delta_{k}X)_{i}) / \Sigma_{i} w_{i}]^2  \), where \( \overline{w}_{i} = 1 / (\sigma_{\mathrm{stat},i}^2 + s_{\mathrm{X}}^2) \).
We ensure that the systematic error is at minimum \( 0.1 \) dex, and add further systematic uncertainties to lines with problematic uncertainties as described in Section \ref{sec:comment}.
This approach provides us a conservative estimate of systematic uncertainties from other sources (e.g., atomic data uncertainties, 1D model atmospheres, and the LTE assumption), given it overexplains some line-to-line variance from stellar parameters in calculating \( s_{X} \).

\begin{figure*}[t]
	\centering
	\includegraphics[width=\textwidth]{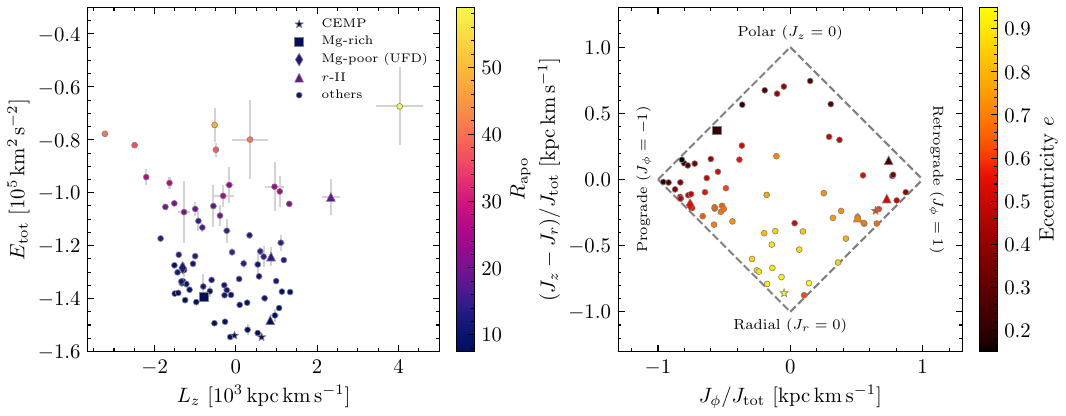}
	\caption{Orbital and kinematic properties for our stars. Left: total energy versus angular momentum, colored by apastron distance \( R_{\mathrm{apo}} \). Right: action space, \( (J_{z} - J_{r}) / J_{\mathrm{tot}}\) against angular momentum fraction \( J_{\phi} / J_{\mathrm{tot}} \). Our orbital properties are consistent with expectations for halo VMP stars. CEMP stars, \( r \)-process enhanced stars, and magnesium unusual stars are highlighted per the legend.}
	\label{fig:kinematics}
\end{figure*}

\begin{figure*}[t]
	\centering
	\includegraphics[width=0.95\textwidth]{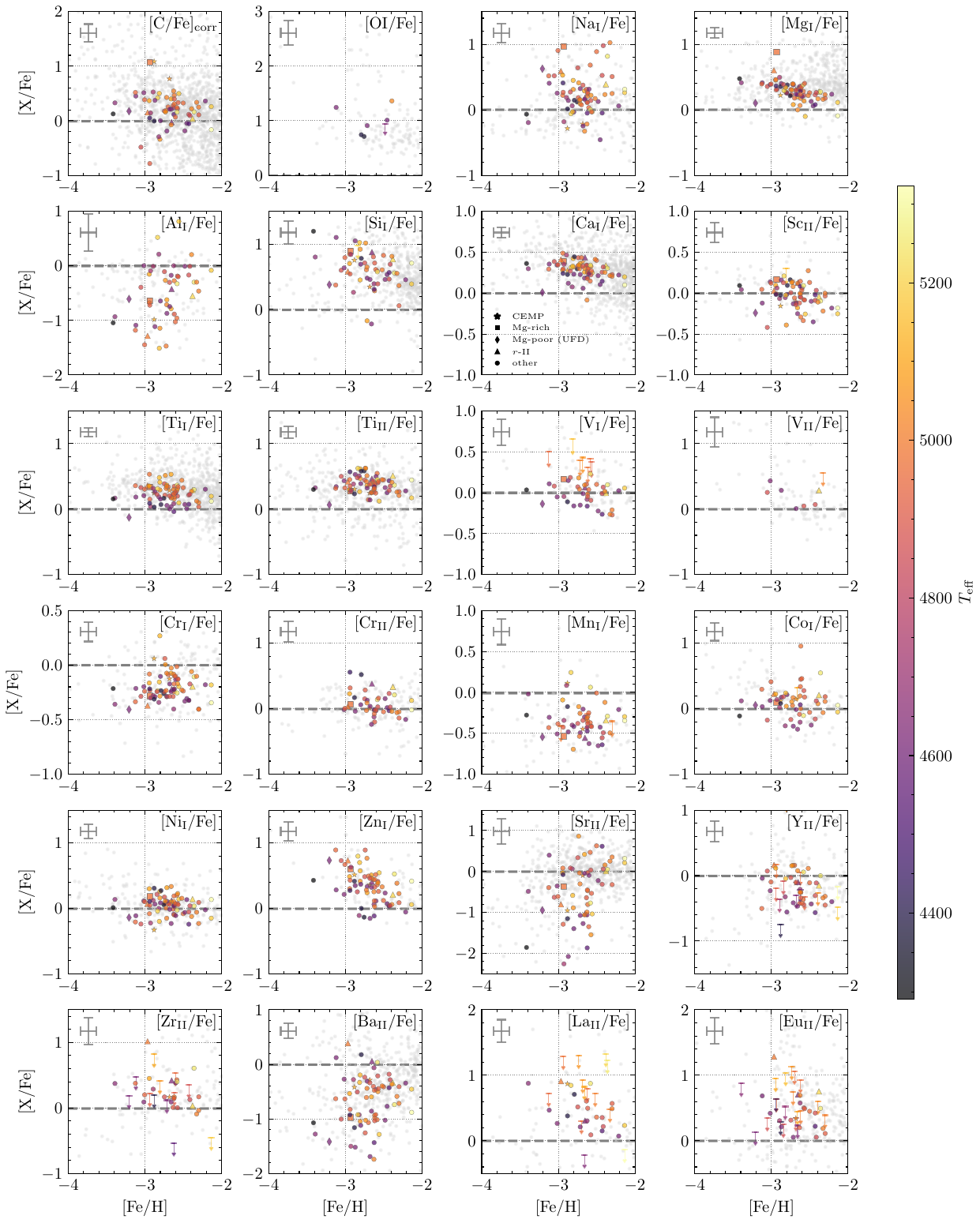}
	\caption{Chemical abundances and upper limits \( \mathrm{[X / Fe]} \) against metallicity \( \feh \) for our 21 elements (24 species) across our 75 metal-poor stars. Points are colored by effective temperature \(\teff\), with chemically unusual stars highlighted and median errors shown in the top left of each panel. Abundances of halo stars from the Stellar Abundances for Galactic Archaeology (SAGA) database are underplotted in gray \citep{sudaStellarAbundancesGalactic2008}. \( \mathrm{[C / Fe]} \) is corrected for evolutionary effects based on \citet{placcoCARBONENHANCEDMETALPOORSTAR2014}. The chemical abundances we derive are consistent with expected ranges for Milky Way halo stars. The complete table of chemical abundances will be made available upon publication.}
	\label{fig:abundances}
\end{figure*}

Our chemical abundance measurements and upper limits for all measured elements are presented in Figure \ref{fig:abundances} and Table \ref{tab:abund}, including relevant ionization states where measured. All quoted \( \mathrm{[X / Fe]} \) measurements are against \( \mathrm{Fe_{I}} \).
There are total of 13 (17.3\%) extremely metal-poor stars (\( \feh \le  -3.0 \)), and the remaining 62 are all very metal-poor stars (\( -3.0 < \feh \le  -2.0 \)). Our most metal-poor star is \( \mathrm{[Fe / H]} = -3.43\, \pm\, 0.10 \) dex. If consider our removed OBA-type contaminant stars, the purity of our metal-poor selection is 75/90 (\( 83.3\% \)), and the selection purity in the disk plane (\( |b| \le  10 \)) is 6/21 (\( 28.6\% \)). All stars outside the plane of the disk are metal-poor stars. The contamination of OBA-type stars along the disk plane is expected, given their spectroscopic similarity at with high reddening.
Of our sample, 20 very metal-poor stars and 2 extremely metal-poor stars have not been identified as metal-poor candidates previously. While three of our stars were previously identified as spectroscopic binaries \citep{gaiacollaborationGaiaDataRelease2023a}, we do not find any features of binarity in their spectra.
Specific comments on the abundance measurements are given in Section \ref{sec:comment}.

\subsection{Kinematics analysis}\label{sec:kinematics}
\begin{figure*}[t]
	\centering
	\includegraphics[width=\textwidth]{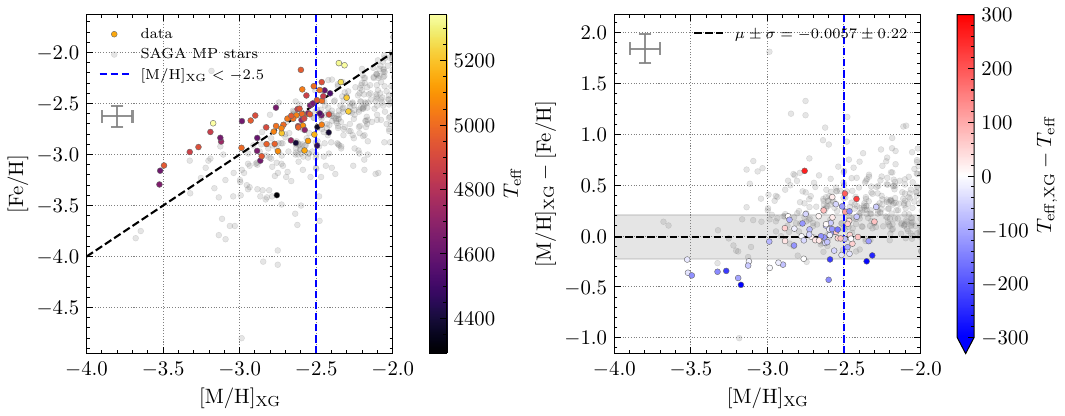}
	\caption{The original \texttt{XGBoost} selection remain accurate, even down to low metallicities. Shown is a comparison of \( \mh\) estimates and \( \feh \) measurements, colored by \( \teff \). Left: direct comparison; \( \feh \) from our work versus \( \mh \) estimates from \citet{andraeRobustDatadrivenMetallicities2023}. Right: differences between estimations (XG - our results). The selection cut applied to the catalog for observations is overplotted in blue. Both plots show minimal scatter for the majority of the stars, with outliers driven by differences in effective temperature. Metal-poor stars from the SAGA database are also underplotted, which are consistently labelled as metal-poor stars.}
	\label{fig:mh}
\end{figure*}

For all of our stars, we utilize the 6D phase-space information to calculate their orbital parameters. Orbits and their properties are computed following the approach of \citet{rixPoorOldHeart2022}.

We adopt a four-component Milky Way mass, consisting of a spherical Hernquist nucleus and bulge component, an (approximate) exponential disk component, and a spherical Navvaro-Frenk-White halo component. The potential model is fit to the \citet{eilersCircularVelocityCurve2019} rotation curve and a compilation of Milky Way enclosed mass measurements \citep[Section 2,][]{huntMultiplePhaseSpirals2022}, as implemented in the \texttt{MilkyWayPotential2022} class in \texttt{Gala}\footnote{\href{https://github.com/adrn/gala}{github.com/adrn/gala}} \citep{price-whelanAdrnGalaV1912024}. Orbits are integrated using the \texttt{Gala} code for a timestep of 0.5 Myr for a total of 2 Gyr. To solve for actions, we adopt the Stäckel-Fudge method \citep{binneyMoreDynamicalModels2012, sandersReviewActionEstimation2016} for this axisymmetric potential of the Galaxy with \texttt{GalPy}\footnote{\href{https://github.com/jobovy/galpy}{github.com/jobovy/galpy}} \citep{bovyGalpyPythonLIBRARY2015}. Radial velocities are obtained from \texttt{PayneEchelle} as described in Section \ref{sec:obs}.

Figure \ref{fig:kinematics} shows our fundamental dynamics parameters for our integrated orbits. The left panel shows the angular momentum against total energy of the orbits, indicating that many of our stars are retrograde, bound stars in the Milky Way. The right panel shows the action space, which is consistent with expectations for VMP halo stars. Chemically unusual stars are highlighted per the legend, and discussed further in Section \ref{sec:outliers}. We see no particular clumping or exclusion in the dynamical space, which is expected given the target selection of mostly halo stars from \emph{Gaia}'s all-sky coverage.

\section{Discussion}\label{sec:discussion}
\subsection{Metallicity estimation validation and comparisons}\label{sec:validate}
\subsubsection{Validation of \texttt{XGBoost} metallicities}

To assess the robustness of the \citet{andraeRobustDatadrivenMetallicities2023} catalog, we compare our measured metallicities \( \feh \) to their initial \( \mh_{\mathrm{XG}} \) estimates from \citet{andraeRobustDatadrivenMetallicities2023} in Figure \ref{fig:mh}. The minimal scatter is particularly encouraging, especially for the region around \( \feh < -3.0 \),  and variations from the one-to-one relation are roughly within \( 1\text{--}2\sigma \). We also see that metallicities of stars from the SAGA database \citep{sudaStellarAbundancesGalactic2014} are also consistent to lower metallicities. The estimations lean towards higher values (below the 1:1 line) than lower values (being above the 1:1 line), such that the model does not lean towards labelling stars as more metal-poor than they truly are. Thus, we conclude that the \texttt{XGBoost} metallicity estimates in \citet{andraeRobustDatadrivenMetallicities2023} are sufficiently robust to motivate follow-up and targeting for the most metal-poor stars.

\subsubsection{Exclusion of more metal-poor candidates}
\begin{figure}[t]
	\centering
	\includegraphics[width=\columnwidth]{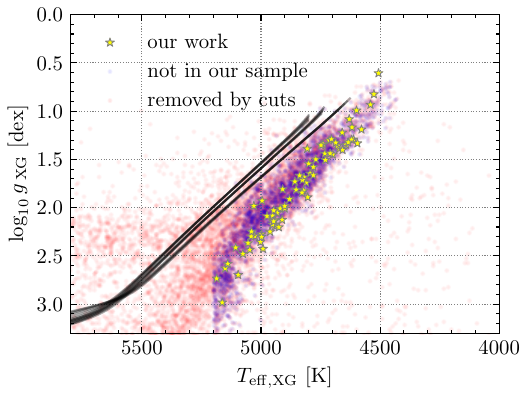}
	\caption{Stars with metallicity \( \mh < -2.5 \) from the RGB sample in \citet{andraeRobustDatadrivenMetallicities2023} and our work compared against the MIST isochrones for \( \feh = -3.0, -3.5, -4.0 \). All stars from the catalog are below the isochrones for \( \mh \le -2.5\), which suggested exclusion of some extremely metal-poor candidates through the cuts used in creating the RGB catalog. We plot the stars removed by the quality cuts we believe could be relaxed to net an increased sample size.}
	\label{fig:andrae_cuts}
\end{figure}

We initially noticed an absence of EMP (\( \feh \le -3.0 \)) stars in our sample along the MIST isochrones, as shown in the Figure \ref{fig:params}. To assess this, we tested the complete vetted RGB catalog in \citet{andraeRobustDatadrivenMetallicities2023}, which present a lack of extremely metal-poor star candidates (blue) close to the theoretical predicted isochrone tracks for metal-poor stars.
We find that this arises from the empirical selection cuts for the vetted RGB sample, as described in Table 2 of \citet[][]{andraeRobustDatadrivenMetallicities2023}:
\begin{enumerate}
	\item \( \sigma (\overline{\omega } ) / \overline{\omega } > 5    \)
	\item \( T_{\mathrm{eff,XG}} < 5200\,\mathrm{K} \)
	\item \( W_{1} > -0.3-0.006 \times (5500 - T_\mathrm{eff, XG}) \)
	\item \( W_{1}> -0.01 \times (5300 - T_\mathrm{eff, XG}) \)
\end{enumerate}
where \( W_{1}\) is the absolute magnitude of stars in the CatWISE W1 photometry band  \citep{maroccoCatWISE2020Catalog2021}, and \( \overline{\omega }  \) is the parallax. The cuts are designed to create a high-quality sample by excluding stars with poor parallax solutions and excluding a region in magnitude and effective temperature space where OBA-type stars typically appear. OBA-type stars are not part of the training set of \citet{andraeRobustDatadrivenMetallicities2023}, as they reside outside the SDSS/APOGEE \( \teff \) range. The stars cut from the vetted RGB catalog by these four criteria are shown as red points in Figure \ref{fig:andrae_cuts}.

These cuts introduce a systematic selection effect -- faint, distant metal-poor giant stars will have both poor parallax solutions and higher surface temperatures. For instance, the new ultra metal-poor (UMP; \( -5 < \feh \le  -4 \)) star discovered by \citet{limbergDiscovery$rmFeSim2025} and \citet{jiNearlyPristineStar2025} is cut from Table 2 of \citet{andraeRobustDatadrivenMetallicities2023} by its poor (negative) parallax solution.

For future follow-up work based on the \citet{andraeRobustDatadrivenMetallicities2023} catalog, we thereby suggest removing the parallax quality cut and relaxing the \( \teff \) and \( W_{1} \) constraints (i.e., by 0.05-0.15 mag and 100-200 K) to maximize the sample size of probable metal-poor candidates. Additionally, using the catalog alongside other catalogs of both photometric and XP-based metallicities can be helpful to determine the most probable metal-poor candidates \citep[e.g.,][]{mardiniStrontiumrichUltrametalpoorStar2024,limbergDiscovery$rmFeSim2025,jiNearlyPristineStar2025}.

An alternative approach to handling OBA-type contaminants could be to avoid the disk field altogether.
At the low XP spectral resolution, highly reddened OBA-type stars and metal-poor stars appear similar.
This produces a classification degeneracy where each can be misclassified as the other.
For instance, \citet{loweRiseMilkyWay2025} serendipitously discovered 12 new EMP stars along the Galactic plane, all of which are mislabelled as more metal-rich stars in both target selection \citep{zhangParameters220Million2023} and in \citet{andraeRobustDatadrivenMetallicities2023}.
To better isolate highly probable metal-poor candidates, one could instead cut the disk field entirely and eliminate the majority of highly reddened sources. Other catalogs of XP-based metallicities have followed this approach \citep[e.g.,][]{yangMetallicities20Million2025}, or instead incorporated hot stars and/or reddening information into their modelling to mitigate these degeneracies \citep[e.g.,][]{khalatyanTransferringSpectroscopicStellar2024,yao200000Candidate2024}. We discuss these approaches further in Section \ref{sec:compare}.

\subsubsection{Comparison with other XP-based catalogs}\label{sec:compare}
\begin{figure*}[t]
	\centering
	\includegraphics[width=\textwidth]{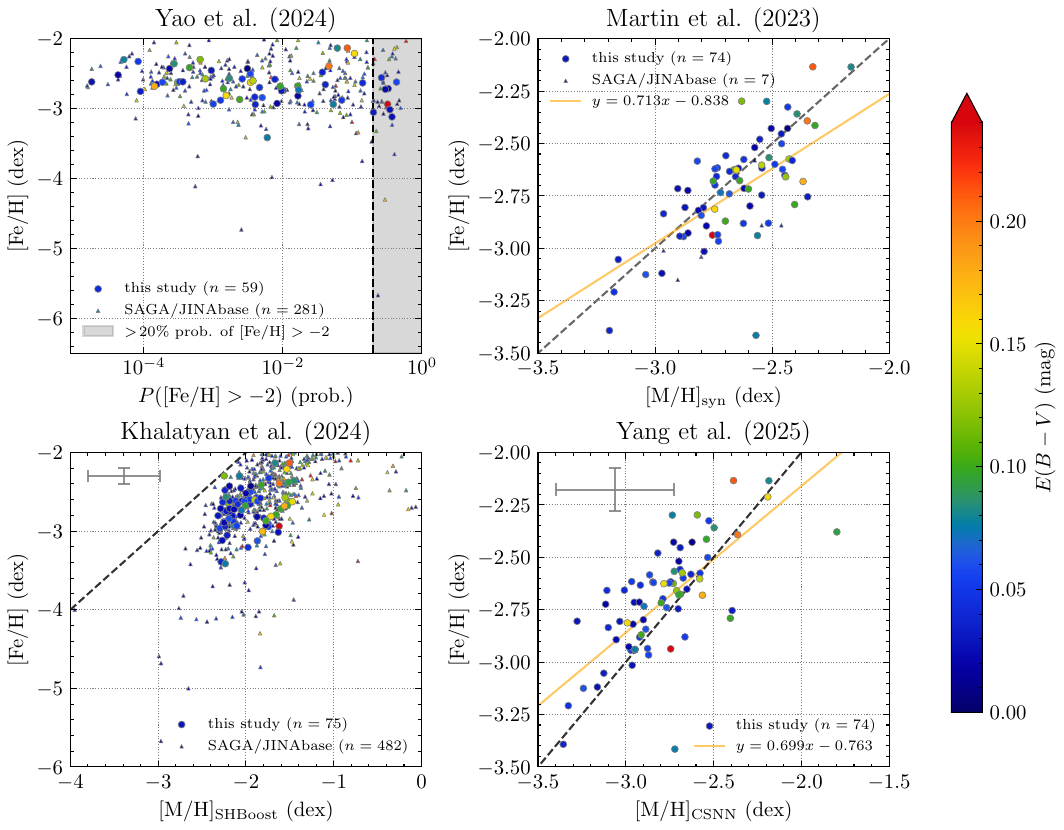}
	\caption{Other catalogs in the literature can robustly model \( \mh \) to low metallicities using different approaches. Comparison of modelled metallicities \( \mh \), augmented with both EMP and UMP stars from the SAGA database to different catalogs of metallicities available in the literature. These are:
		\citet{martinPristineSurveyXXIII2023}, using synthesized CaHK metallicities from Gaia XP spectra.
		\citet{yao200000Candidate2024}, using a dereddened set of XP coefficients and a larger training sample of metal-poor stars. Their probability of being not very metal-poor or lower (\( \feh > -2 \)) is plotted.
		\citet{khalatyanTransferringSpectroscopicStellar2024}, using \texttt{XGBoost} on a dereddened set of Gaia XP coefficients and substantially larger training set that includes hotter stars.
		\citet{yangMetallicities20Million2025}, using a cost-sensitive neural network on large sample of metal-poor stars from the PASTEL/SAGA catalogs.
		Where reported, model errors are plotted in the top left.
		See text for details.}
	\label{fig:compare}
\end{figure*}

Distinguishing metallicities at the \emph{Gaia} XP spectral resolution is complicated by the classification degeneracy between reddened OBA-type stars and metal-poor stars. Parameters which strongly impact the continuum, including metallicity, effective temperature, extinction, and carbon abundance can each be utilized to help break this degeneracy. Various studies have adopted different approaches and combinations of these parameters in their modelling to mitigate this degeneracy. In this Section, we compare the results from some of these works to our spectroscopic sample and compiled metal-poor stars from the SAGA and JINAbase databases \citep{abohalimaJINAbaseDatabaseChemical2018}. All of our comparisons are shown in Figure \ref{fig:compare} and are discussed in detail below.

\citet{yao200000Candidate2024} derive stellar parameters \( \teff \), \( \logg \) and \( \feh \) for nearly 200,000 stars with three different classification models. Compared to \citet{andraeRobustDatadrivenMetallicities2023}, their modelling effort focuses on VMP stars. Their work utilizes the classification version of \texttt{XGBoost}, which means they return probabilities instead of estimated metallicities. Additionally, they explicitly apply reddening corrections to the \emph{Gaia} XP spectra polynomial coefficients by reference to stars of similar stellar parameters.
In Figure \ref{fig:compare}, we plot the probability that the star is \emph{not} very-metal poor (i.e., \( \mathrm{[Fe / H]} > -2 \)).
The \citet{yao200000Candidate2024} catalog is also similarly robust for classifying VMP stars, with the majority of stars having an extremely high probability that \( \feh < -2.0 \). The catalog is also more accurate for classifying UMP stars. This improvement likely comes from their stronger representation in the training set, but we note that this improvement comes with the limitation of being less informative about the true metallicity of the candidate stars. None of our stars have a probability \( P < 0.44 \), but 8 of our targets and 33 SAGA/JINAbase stars have \( P >  0.2 \), and do not show a clear correlation with key properties, including extinction, carbon abundance, stellar parameters, distance, or sky position. We note that the 15 OBA-type stars eliminated from our analysis do not appear in any of their metal-poor candidate catalogs (classifiers GP, GC, T), which is likely a consequence of their reddening corrections to the \emph{Gaia} XP spectra coefficients.

\citet{martinPristineSurveyXXIII2023} do not take a data-driven approach, but instead derive photometric metallicities from the \emph{Gaia} XP spectra by synthesizing a magnitude across the Ca HK lines for the entire \emph{Gaia} DR3 XP spectra catalog. These are then used to estimate \( \feh \). The minimal scatter is encouraging, especially in the \( \mh  < -3 \) region. The minor outliers off the one-to-one line do not appear to show any correlation with the stellar parameters, photometry, nor carbon abundance. Future works could look to compare tradeoffs in accuracy between  synthesized photometry and other data-driven approaches.

\begin{figure}[t]
	\centering
	\includegraphics[width=\columnwidth]{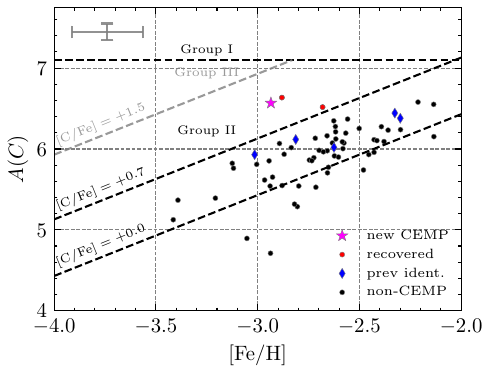}
	\caption{Absolute carbon abundance \( A(C) \) against metallicity \( \feh \), corrected for evolutionary effects from corrections by \citet{placcoCARBONENHANCEDMETALPOORSTAR2014}. Black dashed lines mark the classification of different levels of carbon enhancement according to \citet{yoonObservationalConstraintsFirststar2016, yoonOriginCEMPnoGroup2019}. Previously identified CEMP stars which are part of our targets are highlighted. A new CEMP star is marked in magenta.}
	\label{fig:yoonbeers}
\end{figure}

\citet{khalatyanTransferringSpectroscopicStellar2024} also use a similar methodology to \citet{andraeRobustDatadrivenMetallicities2023}. Their work also uses \texttt{XGBoost}, but they include a larger training set of spectroscopic labels (700K vs 7M) and include hotter stars in their training data. Their work also does not synthesize photometry from the \emph{Gaia} XP spectra in their training process, but they include dereddened photometry and \emph{Gaia} XP spectra from the \texttt{StarHorse} code. We confirm that their metallicity estimations show no strong correlation with \( E(B-V) \). We confirm their finding that \( \mathrm{[M / H]_{SHBoost}} \) is systematically overestimated for all metal-poor stars across both our sample and metal-poor stars from SAGA and JINAbase, but all of the most metal-poor stars are labelled as metal-poor. We find that metallicity can be estimated upwards of 2 dex higher than the known \( \feh \), which may make holistic selections of metal-poor stars challenging given the magnitude of these differences. However, we also find that all 15 of our OBA-type contaminants have well-estimated \( \teff \) and \( \mh \) values in the ranges expected for OBA-type stars (i.e., \( \mh \gtrapprox -1  \)), which may suggest their catalog has less contamination from these stars than other works.

\citet{yangMetallicities20Million2025} use a cost-sensitive neural network, combining both the PASTEL \citep{soubiranPASTELCatalogueStellar2010,soubiranPASTELCatalogue20162016} and SAGA catalogs. For this reason, we do not overplot the SAGA stars for this catalog. We find that the metallicity estimations are similarly well-estimated to the other robust works and \citet{andraeRobustDatadrivenMetallicities2023}, and well within their provided uncertainty.
The quality of estimations in our sample does not appear to be correlated with extinction, sky position, stellar parameters, nor \( \mathrm{[C / Fe]} \).

In summary, we find that most \emph{Gaia} XP-based metallicities are reliably estimated across a range of approaches, whether using data-driven methods and/or synthetic photometry. We find that the majority of metal-poor stars are well-classified by all approaches, and accounting for extinction or removing contaminant regions appears to be successful in reducing the presence of OBA-type contaminants.

\subsection{Interpretation of chemical abundance results}\label{sec:comment}
\begin{figure}[t]
	\centering
	\includegraphics[width=\columnwidth]{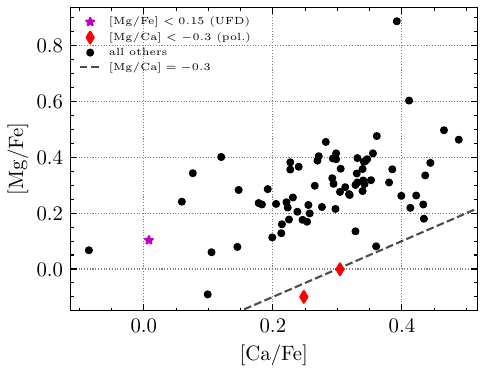}
	\caption{Magnesium-to-iron ratio \( \mathrm{[Mg / Fe]} \) against calcium-to-iron ratio \( \mathrm{[ Ca / Fe]} \). Mg-poor stars are identified with colored points. The two red Mg-poor stars present \( \mathrm{[Mg / Ca]} \lesssim -0.3 \) dex and \( \mathrm{[Na / Fe]} < 0\,\mathrm{dex} \), which is consistent with pollution of their star forming ISM from Type Ia supernovae \citep{ivansChemicalSubstructureMilky2003}. The low occurrence of these signatures (2/75) is expected given how signatures of individual pollution events smear out at \( \feh \sim -2\,\mathrm{dex} \) \citep{hansenRprocessAllianceFirst2018}. The magenta point marks an Mg-poor star which shows dynamical traces of being a part of an accreted system and abundances consistent with an ultra-faint dwarf galaxy \citep{letarteHighresolutionVLTFLAMES2010,frebelSegue1Unevolved2014,jiChemicalAbundancesUltrafaint2019,andalesOldestStarsLow2024}.}
	\label{fig:mgpoor}
\end{figure}

In this section, we provide detailed comments on our chemical abundance results, such as how abundances were derived and relevant caveats.

\subsubsection{Light elements}

\begin{deluxetable*}{rlrrrrrrrrrr}[t]
	\tablecolumns{12}
	\tabletypesize{\scriptsize}
	\tablecaption{\label{tab:abund} Chemical abundances.}
	\tablehead{Gaia DR3 ID & Element  & N & \( \log \epsilon \) (dex) & Lim.  & \( \mathrm{[X/H]} \)  (dex) & \( \mathrm{[X/Fe]} \) (dex) & \( \Delta_{\teff}\rm{X} \) & \( \Delta_{\logg}\rm{X}  \) & \( \Delta_{\vt}\rm{X} \) & \( \Delta_{\mh}\rm{X} \) & \( s_{\mathrm{X}} \) }
	\startdata
	4769570635432428032 & Na I   & 2  & $3.26 \pm 0.01$& N  & $-2.98 \pm 0.21$ & $-0.33 \pm 0.15$ & $0.06$  & $-0.01$ & $-0.19$ & $-0.01$  & $0.10$ \\
	4769570635432428032 & Mg I   & 4  & $4.94 \pm 0.13$& N  & $-2.66 \pm 0.13$ & $-0.00 \pm 0.08$ & $0.06$  & $-0.01$ & $-0.08$ & $-0.00$  & $0.15$ \\
	4769570635432428032 & Ca I   & 13 & $3.99 \pm 0.11$& N  & $-2.35 \pm 0.07$ & $0.30  \pm 0.06$ & $0.04$  & $-0.00$ & $-0.04$ & $-0.00$  & $0.10$ \\
	4769570635432428032 & Sc II  & 2  & $0.61 \pm 0.19$& N  & $-2.54 \pm 0.15$ & $0.12  \pm 0.18$ & $-0.0$2 & $0.03 $ & $-0.01$ & $0.02 $  & $0.15$ \\
	4769570635432428032 & Ti I   & 12 & $2.46 \pm 0.12$& N  & $-2.49 \pm 0.09$ & $0.17  \pm 0.06$ & $0.08$  & $-0.00$ & $-0.02$ & $-0.00$  & $0.10$ \\
	4769570635432428032 & Ti II  & 22 & $2.54 \pm 0.19$& N  & $-2.41 \pm 0.08$ & $0.25  \pm 0.07$ & $0.03$  & $0.02 $ & $-0.06$ & $0.01 $  & $0.18$ \\
	4769570635432428032 & V I    & 1  & $1.37 \pm 0.00$& N  & $-2.56 \pm 0.15$ & $0.09  \pm 0.18$ & $0.04$  & $-0.00$ & $0.03 $ & $-0.02$  & $0.10$ \\
	4769570635432428032 & Cr I   & 5  & $2.96 \pm 0.22$& N  & $-2.68 \pm 0.12$ & $-0.03 \pm 0.11$ & $0.07$  & $-0.00$ & $-0.02$ & $0.00 $  & $0.20$ \\
	4769570635432428032 & Fe I   & 92 & $4.84 \pm 0.26$& N  & $-2.66 \pm 0.10$ & $0.00  \pm 0.04$ & $0.07$  & $-0.00$ & $-0.07$ & $-0.00$  & $0.24$ \\
	4769570635432428032 & Fe II  & 14 & $4.86 \pm 0.24$& N  & $-2.64 \pm 0.09$ & $0.01  \pm 0.10$ & $0.01$  & $0.02 $ & $-0.05$ & $0.01 $  & $0.24$ \\
	4769570635432428032 & Ni I   & 4  & $3.90 \pm 0.23$& N  & $-2.32 \pm 0.14$ & $0.33  \pm 0.13$ & $0.06$  & $-0.00$ & $-0.04$ & $-0.00$  & $0.24$ \\
	4769570635432428032 & Zn I   & 2  & $2.36 \pm 0.18$& N  & $-2.20 \pm 0.14$ & $0.46  \pm 0.15$ & $0.03$  & $0.01 $ & $-0.02$ & $0.01 $  & $0.17$ \\
	4769570635432428032 & Sr II  & 2  & $-0.84 \pm0.60$& N  & $-3.72 \pm 0.51$ & $-1.06 \pm 0.48$ & $0.05$  & $0.02 $ & $-0.28$ & $0.01 $  & $0.57$ \\
	4769570635432428032 & Y II   & 1  & $-0.37 \pm 0.00$& N  & $-2.58 \pm 0.19$ & $0.08  \pm 0.20$ & $0.02$  & $0.03 $ & $-0.01$ & $0.02 $ & $0.10$ \\
	4769570635432428032 & C-H    & 2  & $5.69\pm 0.05$& N  & $-2.74 \pm 0.16$ & $-0.08 \pm 0.14$ & $0.12$  & $-0.03$ & $0.00 $ & $0.05 $ & $0.10$ \\
	\enddata
	\tablecomments{One of the Mg-poor stars is shown for brevity. The full table will be made available upon publication.}
\end{deluxetable*}

Carbon is measured from spectral synthesis of the CH molecular features at 4313 \AA{} and 4323 \AA{} where each region is treated independently. We assume a carbon isotope ratio \( \mathrm{C}_{12} / \mathrm{C}_{13} \) of 9 in our stars, given that most have evolved past first dredge up. In all Figures containing \( \mathrm{[C / Fe]} \), we have applied evolutionary carbon corrections based on the grids from \citet{placcoCARBONENHANCEDMETALPOORSTAR2014}\footnote{\href{https://vplacco.pythonanywhere.com/}{vplacco.pythonanywhere.com}}.

Oxygen is only measured in 6 stars from the equivalent widths of the forbidden line at 6300 \AA{}.

Aluminium is detected in 54 of our stars across the 3944 and 3961\,\AA{} lines, which are unfortunately unreliable. Both lines are in the bluer spectral orders, where the S/N is much lower, are close to strong hydrogen features which affect continuum placement, and are subject to significant non-LTE (NLTE) effects \citep{mashonkinaInfluenceInelasticCollisions2016,nordlanderNonLTEAluminiumAbundances2017,lindNonLTEAbundanceCorrections2022}. We have added an extra 0.3 dex minimum systematic uncertainty to each Al line to account for significant continuum modelling issues and NLTE effects. We caution against use of our Al abundances, as the uncertainties are large, and may still be underestimated.

Three redder Sc II lines (4669, 5030,5526\,\AA) and at least one of five bluer lines from 4246 to 4415 \AA{} are detected in 62 stars. All lines are synthesized due to the hyperfine structure, and the blending with carbon in the bluer features. We additionally add 0.1 dex minimum systematic uncertainty per Sc line due to the hyperfine structure causing the line abundances to be sensitive to the smoothing kernel.

\subsubsection{\( \alpha  \) elements}

All four of the measured \( \alpha  \) elements Mg, Si, Ca, and Ti are enhanced relative to Fe at a similar level, where the unweighted averages of \mgfe, \sife, \cafe, and \tife{} are \( 0.28 \pm 0.15 \), \( 0.61 \pm 0.28 \), \( 0.28 \pm 0.12 \), and \( 0.23 \pm 0.19 \) dex respectively, which are all within \( 1 \sigma  \) of the expected \( \alpha  \) enhancement of 0.4 dex for halo stars \citep{tinsleyStellarLifetimesAbundance1979, vennStellarChemicalSignatures2004}.

The sodium abundances are almost always determined from Na D doublet features at 5889\,\AA, but the weaker feature at 5688\,\AA{} is detected in 5 stars, and agrees with the Na D lines to 0.1 dex. The Na D lines are generally corrected for NLTE effects by \( -0.4  \) to 0.1 dex in cool metal-poor giant stars \citep{andrievskyNLTEDeterminationSodium2007, lindNonLTECalculationsNeutral2011,amarsiGALAHSurveyNonLTE2020}, but we do not apply these corrections here.

Magnesium is measured with equivalent widths of up to 9 lines, with three lines detected in all sources (5172, 5183, 5528\,\AA), and two additional lines are detected in 68 sources (4571, 4702\,\AA). Our magnesium lines do not reach saturation, so we continue to use Gaussian profiles across them. We do not apply Mg NLTE corrections to our stars, but they are at most \( \pm 0.1 \) dex for cool metal-poor giant stars \citep{osorioMgLineFormation2016,lindNonLTEAbundanceCorrections2022}.

Si is the least reliable \( \alpha  \) element in metal-poor stars, and is measured from the 3905 and 4102\,\AA{}  Si features. We recover Si measurements in 62 stars. We note that the significant scatter in Si is a reflection of the low measurement precision in the bluer spectral orders and additionally the reliability of the two common features.
Where possible, we use the lines from 5690 to 6000\,\AA{} which also return similar abundances.

Calcium is measured using equivalent widths of up to 25 lines, but at minimum 9 lines. At least 17 lines are detected in 50\% of our stars, and four lines are detected in all stars (5588, 6122, 6162, 6439\,\AA).

Both titanium isotopes are measured using equivalent widths. We consider the Ti II abundances more trustworthy than Ti I, given our sample consists of purely metal-poor giants where Ti I may be significantly affected by NLTE effects \citep{mallinsonTitaniumAbundancesLatetype2024}. Additionally, we recover at least 35 Ti II lines in 80\% of our sample, as opposed to just 17 Ti I lines, which leads to an overall lower scatter in the Ti II abundances.

\subsubsection{Iron-peak elements}
\begin{figure*}[t]
	\centering
	\includegraphics[width=\textwidth]{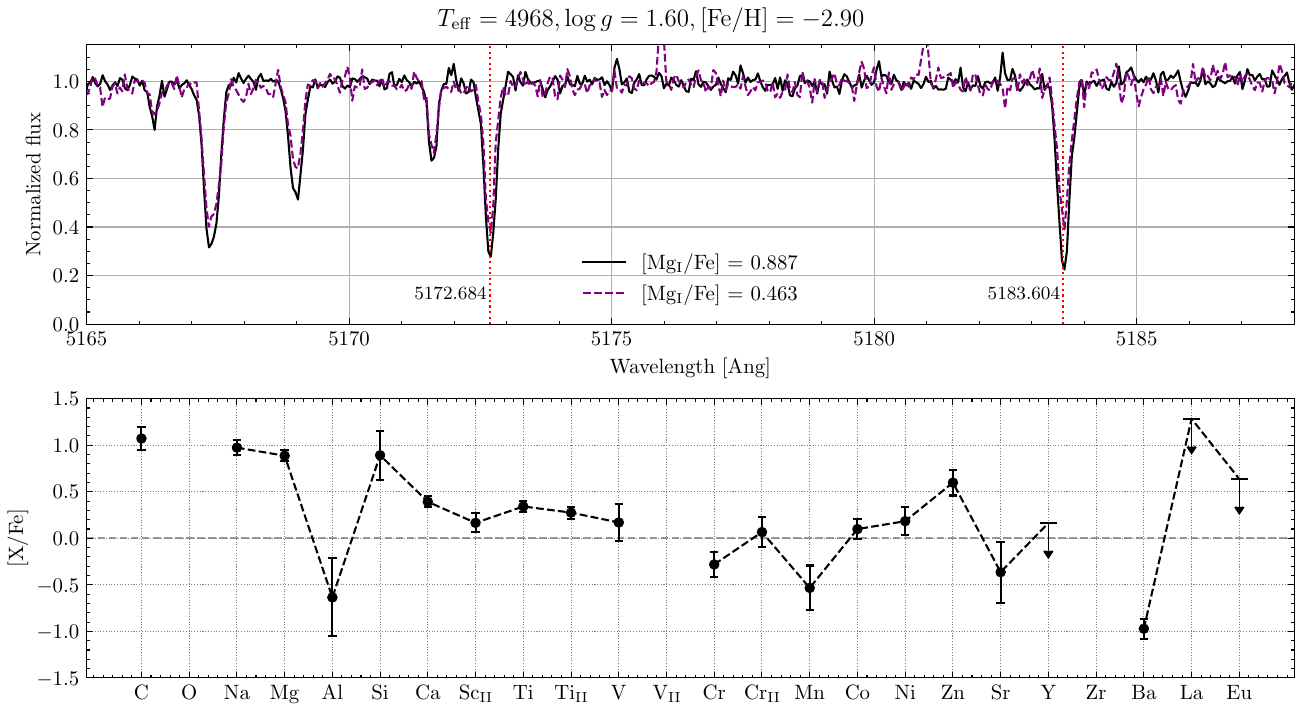}
	\caption{The new CEMP star, Gaia DR3 61674414392593088, also shows enhancement in Mg, Na, and Si, indicative of a progenitor faint CCSNe. Top: Spectrum of Mgb triplet for the star with the strong Mg transitions labelled in red. A comparison star of similar stellar parameters with Mg-normal abundances is plotted in blue. Bottom: Measured elemental abundances.}
	\label{fig:mgrich}
\end{figure*}

For iron, we measure 162 Fe I lines and 29 Fe II lines. The median number of recovered lines is 106 Fe I lines and 16 Fe II lines, with at least 12 Fe II lines are detected in 90\% of our stars. It is difficult to recover lines in the bluer spectral orders due to dark current and the lower S/N.

We use the Fe II lines for determining the microturbulence \( \nu_{\mathrm{t}} \). As we did not balance ionization states, the isotopic abundances differ by \( 0.07\pm 0.06 \) dex. There is one star in our sample where due to the low S/N, we only recover 7 Fe II lines, which adversely affects the microturbulence error (\( u(\nu_{\mathrm{t}}) = 1.3\,\mathrm{km\,s^{-1} } \)) and thereby the Fe I abundances error.

We measure V I at 4384\,\AA{} and 4379\,\AA{} in 48 stars. While we synthesize for 3 lines, we only recover the V II line at 4006\,\AA{} due to the low S/N across the other two lines. Additionally, we only recover V II measurements in 8 stars.

Cr I is measured in all 75 of our stars from 15 lines, and we use at least 11 lines in 80\% of our sample. Cr II is mostly measured across the 4558\,\AA{} and 4824\,\AA{} features, but only in 45 of our stars. We do not correct for Cr I NLTE effects \citep[e.g.,][]{bergemannChromiumNLTEAbundances2010}, so the Cr II abundances should have fewer systematic errors.

Cobalt was measured across the 4020, 4110, 4119 and 4121 \,\AA{}  features. For the 54 stars where Co is detected, it is almost always detected in the 4119 and 4121 \,\AA{} features.

Ni I is measured from equivalent widths of up to 27 lines, however the median number of detected lines is 5. At most, we measure from 17 lines, with at least 9 lines recovered in 80\% of our stars. The 5476\,\AA{} line is always detected, and the next strongest lines are 5137, 4714, and 5080\,\AA{}.

Zn I is measured at 4810 and 4722\,\AA{} using equivalent widths. At least one is detected in 71 of our stars, and both are detected in 39 of our stars.

Seven Mn lines are synthesized, with at least 1 detected in 58 of our stars. We add an additional \( 0.1  \) dex uncertainty to the 4762 and 4766\,\AA{} features, as these were occasionally affected by dark current in the nearby continuum to the absorption features, but not across the features themselves.
While manganese NLTE corrections are available in the literature \citep{bergemannNonLTELineFormation2012}, since our observed abundances fall within the ranges expected for LTE Mn abundances in halo stars, we do not apply any non-LTE corrections to them. Like Sc, we also add an additional 0.1 dex systematic uncertainty (total 0.2 dex) because the hyperfine structure causes these lines to be sensitive to the smoothing kernel.

\subsubsection{Heavy elements}
\begin{figure*}[t]
	\centering
	\includegraphics[width=\textwidth]{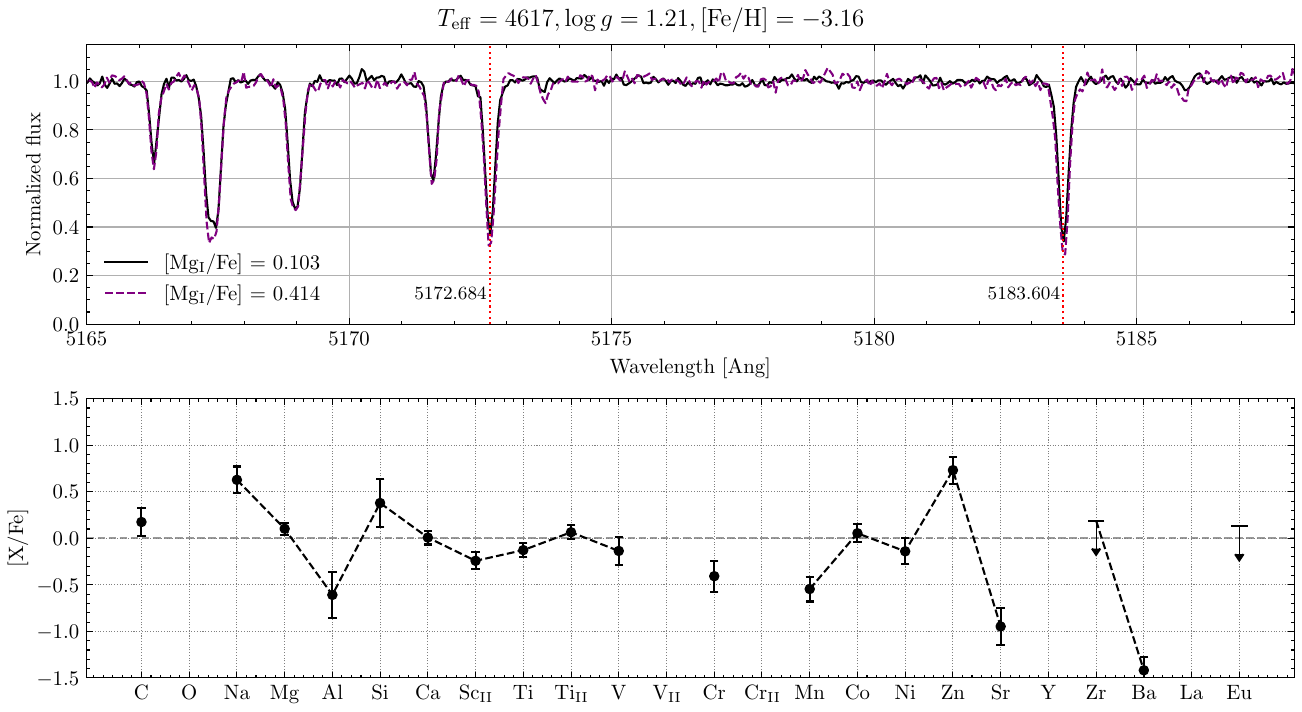}
	\caption{The chemical abundances of Mg-poor star Gaia DR3 5731383034718059520 are indicative of origin from a small accreted system or ultra-faint dwarf galaxy. Minimal enhancement in \( \alpha  \) elements and large depletion in \( r \)-process elements (Sr, Ba, Eu) are visible. Top: Spectrum of Mgb triplet for the star (\( \feh -3.16\, \pm \, 0.09 \)), with the strong Mg transitions marked in red. A comparison star of similar stellar parameters with Mg-normal abundance is plotted in blue. Bottom: Measured elemental abundances.}
	\label{fig:chartmgpoor}
\end{figure*}

At least one Sr II line is measured across 64 stars. The features at 4078\,\AA{} and 4215\,\AA{} are measured using equivalent widths, but we use spectral synthesis for stars where the 4215\,\AA{} feature could be blended with CN molecular features (i.e., \( \mathrm{[C / Fe]} > 0 \) dex). In some stars, the lines are saturated. The propagated microturbulence error dominates the Sr II abundance error for most stars. When detected, Y and Zr abundances are better tracers of a similar nucleosynthetic process.

Yttrium is measured by spectral synthesis for 48 stars across the 4387, 4884, and 4900\,\AA{} Y II features. If all of these are detected, we additionally use up to five other Y II lines at 3950, 4855, 5087, 5206\,\AA{}. These lines are only used in 8 stars.

A single Zr II line is measured by spectral synthesis or upper limit at 4208\,\AA{} in 37 stars. We note these may be weak limits, given the metal-poor nature of our sample.

Barium is measured using spectral synthesis in almost all of our stars (71) across at least 2 lines. We detect one of the strong 4554 or 4933\,\AA{} lines in at least 70 of our stars, and additionally detect at least one of the 5854, 6142, and 6496\,\AA{} features in most stars. We use \( r \)-process isotope ratios for all stars for the synthesis \citep{snedenNeutronCaptureElementsEarly2008}, but we note that using solar isotopic ratios can increase these abundances by up to 0.25 dex. To account for this and the sensitivity to the smoothing kernel from hyperfine splitting, we add an extra systematic uncertainty of 0.1 dex to the barium line abundances.

Lanthanum is measured with spectral synthesis, and is detected in only 24 of our stars. We consider six La features, but primarily detect La from 4922\,\AA{} and 4122\,\AA{}. We return upper limits where possible from the 4086\,\AA{} line, though these are weak limits. These features are blended with carbon and are sensitive to the synthesis smoothing kernel, so we add an additional 0.1 dex systematic uncertainty.

Europium is measured from the Eu II 4129\,\AA{} absorption feature or the Eu II 4205\,\AA{} feature. We found that there was a significant effect from high dark current across the Eu II 4129\,\AA{} feature and occasionally the 4205\,\AA{} feature, which prevented us from obtained accurate europium measurements in 20 stars. Stars with dark current are reported in the expanded version of Table \ref{tab:params}. For consistency, we do not provide upper limits nor abundances for spectra with significant sloping across the continuum from nearby high dark current. Further information on the dark current with the FEROS spectrograph is given in Appendix \ref{sec:darkcurrent}.

\subsection{Chemically unusual stars}\label{sec:outliers}
\subsubsection{Carbon-enhanced metal-poor (CEMP) stars}
Carbon-enhanced metal-poor (CEMP) stars are metal-poor stars that show an abnormal enhancement of carbon relative to iron content (\( \mathrm{[C / Fe]} > +0.7 \)), which provide unique insights into the earliest supernovae of the Milky Way \citep{beersDiscoveryAnalysisVery2005,norrisMostMetalpoorStars2013,frebelNearFieldCosmologyExtremely2015}. Different types of CEMP stars have been identified by \citet{beersDiscoveryAnalysisVery2005}, where the three main classes are CEMP-s stars (\( s \)-process enhancement), CEMP-r stars (\( r \)-process enhancement) and CEMP-no stars (no heavy element enhancement). Carbon excess in CEMP-no stars is attributed to nucleosynthetic pathways associated with the first stars in the Galaxy \citep{iwamotoFirstChemicalEnrichment2005,meynetEarlyStarGenerations2006}, whereas carbon excess in CEMP-s stars is considered to be from mass transfer from an AGB star companion given their radial velocity variations and overabundance of Ba \citep[e.g.,][]{starkenburgBinarityCarbonenhancedMetalpoor2014,placcoHubbleSpaceTelescope2015,hansenRoleBinariesEnrichment2016,hansenRoleBinariesEnrichment2016a,jorissenBinaryPropertiesCH2016}. CEMP-r stars are expected to be enriched from stochastic supernovae events, and are discussed in Section \ref{sec:rp}.

Our stars are shown in the Yoon-Beers \( A(C) \) diagram in Figure \ref{fig:yoonbeers}. At the VMP regime and below, CEMP stars are expected to comprise 20-40\% of a given selection \citep[e.g.,][]{yongMostMetalpoorStars2013,leeCarbonenhancedMetalpoorStars2013,placcoCARBONENHANCEDMETALPOORSTAR2014}.
For the 64 stars we measure carbon in, we find a total occurrence rate of 4/64 ($6.25$\%). It is likely that the \emph{Gaia} XP-based selections preferentially select holistically metal-poor stars, i.e., stars with a low \( \mathrm{[M / H]} \) (including low \( \mathrm{[C /H]} \)) and not solely just a low \( \mathrm{[Fe / H]} \). Similar selection effects are found in photometric selections \citep[e.g.,][]{martinPristineSurveyXXIII2023,huangSpectroscopyIIStellar2023,hongCandidateMembersVMP2024,luStellarLociVII2024} and other \emph{Gaia} XP-based selections \citep{yangMetallicities20Million2025}.

In our observed sample, we find one new CEMP star, TYC 7271-910-1 (Gaia DR3 6167441439259305088), with \( \mathrm{[C / Fe]_{\mathrm{corr}}} = 1.07\, \pm\, 0.12 \) dex. All our new and recovered CEMP stars lie in the Group II classification taxonomy. Group II stars are mostly CEMP-no stars according to \citet{yoonObservationalConstraintsFirststar2016}, although some stars do show enhancement in \( s \)-process or \( r \)-process elements. The new CEMP star does not show any particular enhancement in barium (\( \mathrm{[ Ba / Fe]} < 0.0 \)), and is thus a CEMP-no star. This star is particularly enhanced in magnesium (\( \mathrm{[Mg / Fe]} = 0.86\, \pm\, 0.09 \) dex), and its chemistry is further discussed in Section \ref{sec:mgrich}.

We do not recover 7 CEMP stars that have been previously reported as such by previous works \citep{allenElementalAbundancesClassification2012, arentsenBinarityCEMPnoStars2019, holmbeckRProcessAllianceFourth2020,limbergDynamicallyTaggedGroups2021, shankDynamicallyTaggedGroups2022}. For two of these stars (HE 1311-0131, UCAC2 2370601), we do not return a carbon abundance. For the remaining 5 stars, we find that our measured carbon abundance is typically \( 0.2\text{--}0.6 \) dex lower. We consider our measurements more accurate, given our high-resolution study should provide a more accurate continuum placement than the previous low-resolution studies which classified these stars \citep{limbergDynamicallyTaggedGroups2021,shankDynamicallyTaggedGroups2022}. Two of these 5 stars (Gaia DR3 3520836313191203584, Gaia DR3 5502084317152803200) were also studied in a high-resolution analysis by \citet{holmbeckRProcessAllianceFourth2020}. We can reproduce their reported \( \mathrm{[C / Fe]_{\mathrm{corr}}} \) values with their stellar parameters.

CEMP-no stars dominate the fraction of stars with \( \feh < -3 \) \citep[e.g.,][]{aokiCarbonenhancedMetalpoorStars2007,yoonObservationalConstraintsFirststar2016}. These stars are regarded as direct descendants of Population III stars due to their low metallicity and low abundances of neutron-capture elements \citep[e.g.,][]{christliebStellarRelicEarly2002,frebelNucleosyntheticSignaturesFirst2005,caffauExtremelyPrimitiveStar2011,kellerSingleLowenergyIronpoor2014,yoonObservationalConstraintsFirststar2016,clarksonPopIIIIprocess2018,starkenburgPristineSurveyIV2018,kieltyPristineSurveyXII2021,zepedaChemodynamicallyTaggedGroups2023,lucchesiExtremelyMetalpoorStars2024}. The literature has mainly considered two main channels for CEMP-no formation: the faint supernovae of Population III stars that can eject less iron due to mixing and fallback from their small explosion energy \citep[e.g.,][]{umedaFirstgenerationBlackholeformingSupernovae2003, tominagaSupernovaNucleosynthesisPopulation2007a,hegerNucleosynthesisEvolutionMassive2010,nomotoNucleosynthesisStarsChemical2013,tominagaAbundanceProfilingExtremely2014,ezzeddineEvidenceAsphericalPopulation2019}, or rapidly rotating UMP stars which could pollute the ISM through stellar winds across a carbon-rich surface \citep[spinstars; e.g.,][]{meynetEarlyStarGenerations2006, hirschiVeryLowmetallicityMassive2007,maederFirstStarsCEMPno2015, frischknechtSprocessProductionRotating2016}; see \citet{norrisMostMetalpoorStars2013} for a detailed review of formation mechanisms.

\subsubsection{Mg-enhanced and Mg-poor stars}\label{sec:mgrich}

Magnesium (Mg) abundances in metal-poor stars act as a powerful tracer of the progenitor supernovae.
Being mostly produced from core-collapse supernovae of massive stars \citep{weinbergChemicalCartographyAPOGEE2019,kobayashiOriginElementsCarbon2020}, magnesium has a unique utility to distinguish both the initial mass of progenitor core collapse supernovae (CCSNe) \citep[e.g.][]{jiSpectacularNucleosynthesisEarly2024} and its relative contribution with chemical yields of Type Ia supernovae (SNe Ia) through comparison with other \( \alpha  \) elements. Mg-enhanced metal-poor stars \citep[e.g.,][]{norrisExtremelyMetalpoorStars2002,andrievskyNLTEDeterminationSodium2007, frebelHE13272326Unevolved2008,aokiLAMOSTJ22175059+2104372New2018, kieltyPristineSurveyXII2021} and Mg-poor metal-poor stars \citep[e.g.,][]{ivansChemicalSubstructureMilky2003,vennStellarChemicalSignatures2004,tolstoyStarFormationHistoriesAbundances2009,aokiHighresolutionSpectroscopyExtremely2013,jeongSearchExtremelyMetalpoor2023} are thereby important for observing signatures of supernovae events in their birth environments through their unique chemical signatures.

We find that our new CEMP star, Gaia DR3 6167441439259305088, is also Mg-enhanced, with \( \mathrm{[ Mg / Fe]} = 0.89 \pm 0.06 \) dex. The individual abundances of the star are shown in Figure \ref{fig:mgrich}. It has a metallicity \( \feh = -2.93 \pm 0.09 \) with high carbon enhancement (\( \mathrm{[C / Fe]_{\mathrm{corr}}} = 1.07 \pm 0.12 \) dex). The star is a CEMP-no star, with \( \mathrm{[Ba / Fe]} = -0.97 \pm 0.11 \) dex and upper limit \( \mathrm{[ Eu / Fe]} < 0.64 \). The star is additionally enhanced in sodium with \( \mathrm{[Na / Fe ]} = 0.96 \pm 0.08 \) dex, but not other \( \alpha  \)-elements (e.g., \( \mathrm{[Ca / Fe]} = 0.39 \pm 0.06 \) dex. This enhancement is expected for CEMP-no stars, which can be enriched from a faint CCSNe of a massive progenitor star \citep{nomotoNucleosynthesisStarsChemical2013,frebelNearFieldCosmologyExtremely2015}.

The individual abundance chart and Mgb triplet of this star is shown in Figure \ref{fig:mgrich}. The star demonstrates an enhancement in \( \alpha  \) and light elements including C, Na, Mg, and Si. Additionally, the star shows a relative depletion in the \( s \)-process tracers Sr and Ba. Iron-peak elements remain relatively consistent with the solar ratios.

We find two stars with \( \mathrm{[ Mg / Ca]} < -0.3 \) dex, which suggest the pollution of the star forming ISM from Type Ia supernovae compared to other halo objects of similar metallicities. Magnesium is primarily produced by massive stars, whereas Ca is created by both SNe Ia and CCSNe, which results in a deficiency of Mg relative to Ca from the pollution (i.e., \( \mathrm{[Mg / Ca]} < -0.3 \)) \citep{ivansChemicalSubstructureMilky2003}. The \( \mathrm{[Mg / Fe]} \) ratio of these Mg-poor stars, TYC 7072-1780-1 (Gaia DR3 2895944715088173184) and UCAC4 173-006012 (Gaia DR3 4769570635432428032), is shown in Figure \ref{fig:mgpoor}. These stars present both the low \( \mathrm{[Mg / Ca]} \) and a low \( \mathrm{[Na / Fe]} \) signature indicative of this pollution, with \( \mathrm{[Na / Fe]_{\mathrm{289\ldots }}} = -0.03\, \pm\, 0.16 \,\mathrm{dex}, \mathrm{[Na / Fe]_{\mathrm{4769\ldots }}} = -0.31\, \pm \,0.15 \,\mathrm{dex} \). This gives an occurrence rate for this signature of \( \sim \!2\% \), which is expected as pollution signatures from individual events can smear out around \( \mathrm{[Fe / H]} \sim -2 \,\mathrm{dex} \) \citep{hansenRprocessAllianceFirst2018}.

Another Mg-poor star, TIC 1315034 (Gaia DR3 5731383034718059520), has depletion in Mg (\( \mathrm{[Mg / Fe]} = 0.10 \pm 0.06 \) dex) and is the most \( \alpha  \)-poor star in our sample (\( \afe = -0.01 \) dex).
The abundances and Mgb triplet of this star are shown in Figure \ref{fig:chartmgpoor}.
It is extremely metal-poor (\( \feh = -3.21\pm 0.09 \) dex), and has very low neutron capture element abundances (\( \mathrm{[Sr / H]} = -4.13\, \pm 0.23 \) dex, \( \mathrm{[Ba / H]} = -4.63\, \pm 0.13 \) dex). These abundances are consistent with accreted members of small systems and ultra-faint dwarf galaxies \citep{letarteHighresolutionVLTFLAMES2010,frebelSegue1Unevolved2014,jiChemicalAbundancesUltrafaint2019,andalesOldestStarsLow2024}. The star was first identified as an EMP star by \citet{dacostaSkyMapperDR11Search2019}. The star is on a planar orbit as shown in Figure \ref{fig:kinematics} (diamond marker), within the expected action space region for the early \emph{Gaia}-Sequoia merger \citep{myeongEvidenceTwoEarly2019}, and the star is denoted in Figures \ref{fig:kinematics} and \ref{fig:mgpoor} as from an ultra-faint dwarf galaxy (UFD).

\subsubsection{r-process enhanced stars}\label{sec:rp}
\begin{figure}[t]
	\centering
	\includegraphics[width=\columnwidth]{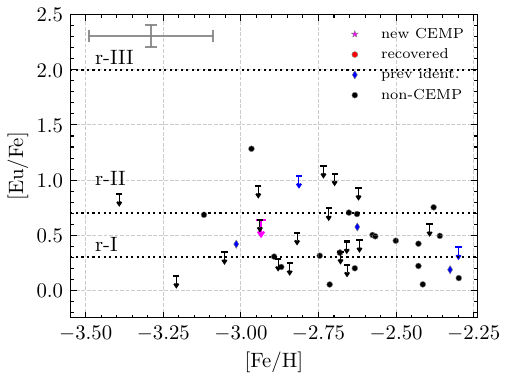}
	\caption{Europium abundance \( \mathrm{[ Eu / Fe]} \) against metallicity \( \feh \), colored by CEMP classification (new CEMP star, recovered CEMP stars from the literature, previously identified CEMP stars, and non-CEMP stars). We recover a total of 15 RPE stars, excluding upper limit candidates. There are 3 stars with measured \( \mathrm{[ Eu / Fe]} > +0.7\) in our sample. All four of these stars have \( \mathrm{[Ba / Eu]}  < 0 \), which classifies them as r-II stars.}
	\label{fig:eufe}
\end{figure}

Stars with \( r \)-process enhancement (RPE) in the chemical abundances of heavy neutron-capture elements (Ba, La, Eu) are important for unraveling and constraining the \( r \)-process nucleosynthesis channels within the Milky Way and its accreted systems \citep[e.g.,][]{hansenRprocessAllianceFirst2018,sakariRProcessAllianceFirst2018,frebelObservationsRProcessStars2023}.

The \( r \)-process classification of our stars is shown in Figure \ref{fig:eufe}, using the classifications by \citet{aokiCarbonenhancedMetalpoorStars2007,holmbeckRProcessAllianceFourth2020}. Our RPE stars are as follows:
\begin{enumerate}
	\item There are a total of 15 RPE stars (\( (\mathrm{[Eu  /Fe]} > 0.3) \wedge (\mathrm{[Ba / Eu]} < 0) \)), excluding stars with upper limits for Eu.
	\item We find 12 r-I enhanced stars (\( 0.3 < \mathrm{[Eu / Fe]} \le  0.7 \)), and 3 r-II stars (\( 0.7 <  \mathrm{[Eu / Fe]}\le 2.0 \))
	\item We recover that one of the r-I stars (BPS CS 22877-0001) is a CEMP-r/s stars (\( \mathrm{[C / Fe]} > 0.7 \) dex), which has been previously reported to be a classical CEMP star \citep{hansenAbundancesKinematicsCarbonenhanced2019,mucciarelliDiscoveryThinLithium2022}.
\end{enumerate}

\section{Conclusion}\label{sec:conclusion}
We have demonstrated the capability of data-driven metallicities derived from XP spectra for finding the most ancient, metal-poor stars in the galaxy. These approaches have been validated through our high-resolution spectroscopic study of 75 red giant branch stars, selected from \citet{andraeRobustDatadrivenMetallicities2023}. We find 20 previously undiscovered very metal-poor stars and 2 new extremely metal-poor stars. We recover a purity of 75/90 in our selection and 6/21 in the disk plane, which is expected the contamination from highly reddened OBA-type stars and the degeneracy for describing the continuum. Our recovered kinematic properties and chemical abundances are consistent with the expected ranges for halo metal-poor stars. We additionally discover a new Mg-enhanced CEMP star and an Mg-poor star from an accreted ultra-faint dwarf galaxy.

Our comparison of the \emph{Gaia} XP metallicity estimations from \citet{andraeRobustDatadrivenMetallicities2023} to our high-resolution results in Figure \ref{fig:mh} shows these estimations are consistent with our \( \feh \) values, even in the extremely metal-poor regime. We additionally highlight the selection effect in the vetted RGB subset presented in \citet{andraeRobustDatadrivenMetallicities2023}, which may exclude further extremely metal-poor star candidates. We further find similarly robust metal-poor classifications in XP-based metallicities presented by \citet{yao200000Candidate2024}, \citet{martinPristineSurveyXXIII2023}, and \citet{yangMetallicities20Million2025}, as shown in Figure \ref{fig:compare}. While the catalog by \citet{khalatyanTransferringSpectroscopicStellar2024} does overestimate metallicities in metal-poor stars, we note their metal-poor candidates may have less contamination from OBA-type stars.

Work on chemodynamical discovery and modelling with the \citet{andraeRobustDatadrivenMetallicities2023} catalog has already been published in the literature \citep[e.g.,][]{chandraThreephaseEvolutionMilky2024, rixExtremelyMetalRich2024, zhangExistenceVeryMetalpoor2024}. Further high-resolution spectroscopic followup based on additional metal-poor candidate star samples has also been published in the literature \citep[e.g.,][]{mardiniStrontiumrichUltrametalpoorStar2024,limbergDiscovery$rmFeSim2025}.

Our work shows the promising scientific utility of \emph{Gaia} XP metallicity catalogs to inform quality target selection of the most metal-poor stars. With the XP spectra in \emph{Gaia} Data Release 4 expected to be of higher quality and to deeper magnitudes (i.e., \( 17.6 < G < 19 \)), it is likely that robust metallicities from \emph{Gaia} XP spectra will continue to be revealed through methods beyond those compared here, each offering a distinct approach to unearth the most ancient stars in the Milky Way.

\section*{Acknowledgements}

A.P.J. acknowledges support from NSF grants AST-2206264, AST-2510795, and an Alfred P. Sloan Research Fellowship.

This work is based on observations made with ESO Telescopes at the La Silla Paranal Observatory under programme ID 112.265J.001.

This work has made use of data from the European Space Agency (ESA) mission Gaia (\href{https://www.cosmos.esa.int/gaia}{cosmos.esa.int/gaia}), processed by the Gaia Data Processing and Analysis Consortium (DPAC, \href{https://www.cosmos.esa.int/web/gaia/dpac/consortium}{cosmos.esa.int/web/gaia/dpac/consortium}).
Funding for the DPAC has been provided by national institutions, in particular the institutions participating in the Gaia Multilateral Agreement. The Gaia archive website is \href{https://archives.esac.esa.int/gaia}{archives.esac.esa.int/gaia}.
This research additionally makes use of public auxiliary data provided by ESA/Gaia/DPAC/CU5 and prepared by Carine Babusiaux.

This research has additionally made use of the NASA Astrophysics Data System (ADS) bibliographic services.

This research has made use of the SIMBAD database, CDS, Strasbourg Astronomical Observatory, France.

This research has made use of the VizieR catalogue access tool, CDS, Strasbourg Astronomical Observatory, France (DOI: \href{https://doi.org/10.26093/cds/vizier}{10.26093/cds/vizier}).

\emph{Software}: \texttt{SIMBAD} \citep{wengerSIMBADAstronomicalDatabase2000}, \texttt{VizieR} \citep{ochsenbeinVizieRDatabaseAstronomical2000}, \texttt{numpy} \citep{harrisArrayProgrammingNumPy2020}, \texttt{scipy} \citep{virtanenSciPy10Fundamental2020}, \texttt{astropy} \citep{astropycollaborationAstropyProjectSustaining2022}, \texttt{pandas} \citep{mckinneyDataStructuresStatistical2010}, \texttt{matplotlib} \citep{hunterMatplotlib2DGraphics2007}, \texttt{vaex} \citep{breddelsVaexBigData2018}, \texttt{dustmaps} \citep{greenDustmapsPythonInterface2018}, \texttt{astroquery} \citep{ginsburgAstroqueryAstronomicalWebquerying2019}, \texttt{ThePayne} \citep{tingPayneSelfconsistentInitio2019}, \texttt{smhr} \citep{caseyTaleTidalTales2014,caseySmhrAutomaticCurveofgrowth2025} and \texttt{LESSPayne} \citep{jiLESSPayneLabelingEchelle2025}.

\bibliography{final}

\appendix

\section{High dark current on the FEROS spectrograph}\label{sec:darkcurrent}
\begin{figure}[htpb]
	\centering
	\includegraphics[width=0.6\columnwidth]{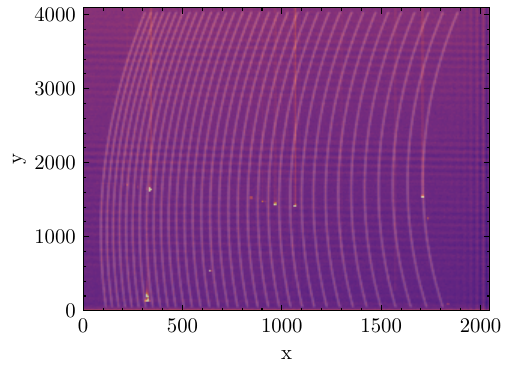}
	\caption{Dark current is visible in the reduced data. An example master bias image overplotted with the echelle spectrograph order traces on the CCD detector for the FEROS spectrograph. Note the vertical streaks visible in orange. While the bias frame shows only an instantaneous read of the electron noise, the structure of the vertical streaks is consistent with dark current in both shape and magnitude. We find that these manifest in the emission-like features at the Eu II absorption line at 4129 \AA{} and across subsequent orders in matching positions along the wavelength solution of our reduced spectra.}
	\label{fig:dark}
\end{figure}
We initially noticed that several europium features showed an emission-like profile in lower S/N observations. This was most common on high magnitude (low brightness) stars on the prominent europium absorption feature Eu II at 4129 \AA{} but we found that it also affected many features between 4000-4500 \AA{} and very rarely features between 4500-4800 \AA{} in low S/N stars.

To assess this, we looked at the intermediate data products output by the CERES pipeline for stars that demonstrated these features. Figure \ref{fig:dark} shows a combined bias master frame overplotted with the trace of the spectrograph. Our bias frame traces show vertical, streak features which cross several echelle traces. The magnitude of this feature is consistent with dark current, and mirrored the flux seen in the output spectrum across different orders. We rule out the possibility of a bad column in the readout as the feature does not persist across the entire image, and is not consistently visible in all images.

During commissioning, the FEROS spectrograph was known to have high dark current issues due to insufficient cooling \citep{kauferCommissioningFEROSNew1999}. These data features are non-recoverable, as there are no dark exposures taken alongside our observations since it is not part of the standard calibration plan for the FEROS spectrograph. The dark current would not affect observations that are of much higher signal-to-noise ratio, but it is relevant for our observations. For features which were consistently impacted by dark current, we have added additional systematic uncertainties or disqualified measurements entirely to ensure a consistent, quality analysis. For instance, we only occasionally measure the Eu II feature at 4129 \AA{} and mostly measure using the Eu II feature at 4205 \AA{} if it is also not skewed by dark current.

Future works for deriving chemical abundances with the FEROS spectrograph at these signal-to-noise ratios (S/N; \( \sim\!10 \to 40 \)) should thus strive to obtain high quality darks alongside their observations.
\end{document}